\documentclass[12pt,journal,draftcls,letterpaper,onecolumn]{IEEEtran}
\usepackage{graphicx, times, amsmath, amsfonts, amsthm, comment}
\usepackage{amssymb}
\usepackage{algorithmic}
\usepackage{algorithm}
\usepackage{color, soul}
\usepackage{mathtools}

\newcommand{\beq}{\begin{equation}}
\newcommand{\eeq}{\end{equation}}
\newcommand{\tbf}{\textbf}
\newcommand{\tit}{\textit}

\newcommand{\argmax}{\operatornamewithlimits{argmax}}
\newcommand{\ud}{\mathrm{d}}

\newcommand*{\mathcolor}{}
\def\mathcolor#1#{\mathcoloraux{#1}}
\newcommand*{\mathcoloraux}[3]{%
 \protect\leavevmode
 \begingroup
  \color#1{#2}#3%
 \endgroup
}

\theoremstyle{plain}
\newtheorem{propcounter}{Proposition}
\newtheorem{proposition}[propcounter]{Proposition}
\theoremstyle{plain}
\newtheorem{lemcounter}{Lemma}
\newtheorem{lemma}[lemcounter]{Lemma}
\theoremstyle{plain}

\theoremstyle{plain}

\theoremstyle{plain}

\newcommand {\Ebb}{\mathbb{E}}

\newcommand {\Ccal}{\mathcal{C}}
\newcommand {\Fcal}{\mathcal{F}}

\newcommand {\Kcal}{\mathcal{K}}

\newcommand {\Scal}{\mathcal{S}}

\begin{document}

\title{Edge Caching in Delay-Constrained Virtualized Cellular Networks: Analysis and Market}

\author{
\IEEEauthorblockN{Tachporn Sanguanpuak\IEEEauthorrefmark{1}, Sudarshan Guruacharya\IEEEauthorrefmark{2}, Ekram Hossain\IEEEauthorrefmark{2} \\
Dusit Niyato\IEEEauthorrefmark{3}, Nandana Rajatheva\IEEEauthorrefmark{1}, Matti Latva-aho\IEEEauthorrefmark{1}}

\thanks{T. Sanguanpuak, N. Rajatheva, and M. Latva-aho are with the Centre for Wireless Communications (CWC), Dept. of Commun. Eng., University of Oulu, Finland (E-mails: \{tachporn.sanguanpuak, nandana.rajathava, matti.latva-aho\}@oulu.fi). S. Guruacharya and E. Hossain are with the Department of Electrical and Computer Eng., University of Manitoba, Canada (E-mails: \{Sudarshan.Guruacharya, Ekram.Hossain\}@umanitoba.ca). D. Niyato is with School of Comp. Science and Eng., Nanyang Technological University (E-mail: dniyato@ntu.edu.sg).}
}

\maketitle

\begin{abstract}
Caching of popular contents at cellular base stations, i.e., edge caching, in order to eliminate duplicate transmission through the backhaul can reduce the latency of data delivery in $5$G networks. However, since caching can only reduce the backhaul delay, techniques such as base station densification will also need to be used to reduce the fronthaul delay. In this paper, using results from stochastic geometry, we first model the effects of base station densification and cache size on the latency of the system. We then derive a tight approximation for the cache hit probability. To optimize the network cost due to the deployment of base station (BS) and cache storage, a minimization problem for the product of the BS intensity and cache size is formulated under probabilistic delay constraint, which is converted into a geometric program and solved analytically. The results are then used to analyze the economics of a cache-enabled virtualized cellular network where the network infrastructure, i.e., BSs and cache storage, owned by an infrastructure provider (InP) is shared among multiple mobile network operators (MNOs). For the pricing between the InP and the MNOs, we formulate a Stackelberg game with the InP as the leader and multiple MNOs as the followers. In this virtualized scenario, the common cost of renting the infrastructure is shared in a fair manner among the MNOs by using the Shapely value. An efficient algorithm is provided to divide the rent among MNOs.

\end{abstract}
\newpage

\section{Introduction} \label{section:introduction}

\subsection{Motivation}
In the upcoming $5$G cellular networks, the traffic volume will be increased by 50 billion devices connected to the cloud by 2020~\cite{Gupta_2015}. Achieving high throughput and very low latency are some of the main challenges in $5$G deployment. In~\cite{Cisco2016}, it was shown that more than $60\%$ percent of the traffic is due to the duplicate downloads of popular video files. The redundancy of data transmission causes severe delay due to backhaul transmission. Therefore, the concept of proactive caching at an edge/radio access network (RAN) or intermediate nodes has be introduced to reduce the backhaul delay~\cite{Wang_commag_2014}. Apart from the backhaul delay, the fronthaul delay also needs to be reduced in order to reduce the overall delay. One possible technique to reduce the fronthaul delay is via BS densification. This has the effect of reducing the average number of user equipments (UEs) served by each BS; thus reducing the fronthaul delay. 

Another important aspect related to $5$G deployment is the question of managing the infrastructure more effectively. One prominent solution is to allow the mobile network operators (MNOs) to share their infrastructure through network virtualization~\cite{Cisco2016, kun-tmc15, Tachporn_TMC2017, Tachporn_ICC2017}. By sharing the infrastructure, e.g., BSs and cache storage, each MNO can maximize the use of the infrastructure while simultaneously minimizing the operational costs. Subsequently, we propose the concept of caching with infrastructure sharing with the constraint on latency for downlink transmission. This scenario presents a new economic model which leads to novel insightful analysis, optimization problem, and game theoretical modeling. 

\subsection{Related Work}
\subsubsection{Caching at RAN/edge caching}
In general, caching in wireless networks consists of two stages: they are cache placement and data delivery. In \cite{Wang_commag_2014}, the authors surveyed various emerging caching strategies for mobile cellular networks and the potential research challenges and opportunities. During the cache placement stage, based on the popularity based caching, the most popular files will be cached in RAN storage or intermediate nodes during off-peak traffic period. During the data delivery stage, the requested files from UEs will be delivered from RAN caching storage or the intermediate nodes. In~\cite{Ejder_2015}, the benefits of popularity-based caching in terms of backhaul offloading and request satisfaction was investigated. A collaborative hierarchical caching at both the cloud RAN and BSs with the aim of minimizing the delay of content delivery under UEs' QoS constraints was proposed in~\cite{Xinhua2017}. In~\cite{ZChen_2016}, a Stackelberg game was used to model the scenario where the BSs reward the UEs for sharing contents with other UEs using device-to-device (D2D) communications. The BS tries to minimize its cost while choosing the rewarding policy, whereas the UE tries to maximize its reward while choosing its caching policy. 

\subsubsection{Economic modeling of caching}
In~\cite{JunLi_2016, JunLi2016a} stochastic geometry was used to characterize the event that a mobile user finds a video file belonging to a given application provider in the cache of a BS closest to it. Using this probability, a Stackelberg game was formulated to jointly maximize the average profit of the network service providers, which act as leaders, and the content providers, which act as followers. In~\cite{Li_2015}, the authors proposed the idea of caching as a service, where the cache service is virtualized. The authors formulate inter MNO and intra MNO traffic load minimization problem. In~\cite{Shen_2016}, the authors considered a Stackelberg game with a single MNO and multiple content providers. The MNO as leader decides the price to charge the content providers such that it maximizes it's revenue. The content providers as followers compete with each other to obtain sufficient cache space to improve the quality-of-service to its users. In~\cite{Zhiwen_2016}, various game theoretical approaches to wireless proactive caching was surveyed. 


In this paper, for a large-scale cache-enabled cellular network, we analyze the tradeoff between BS intensity and the cache storage under delay constraint at a typical UE. For this, we use the downlink SINR coverage probability and throughput obtained based on a stochastic geometry analysis. A tight approximation of cache hit probability in terms of the cache size is also analyzed. Subsequently, we introduce a new economic modeling approach for cache-enabled cellular networks with virtualized BS and cache storage. The pricing problem between the infrastructure provider (InP) and multiple MNOs is formulated using two game theoretical models, namely, a Stackelberg game and a  cooperative game. 

\subsection{Contribution}
\begin{figure}[h]
\begin{minipage}{0.45\textwidth}
\begin{center}
	\includegraphics[width=\textwidth]{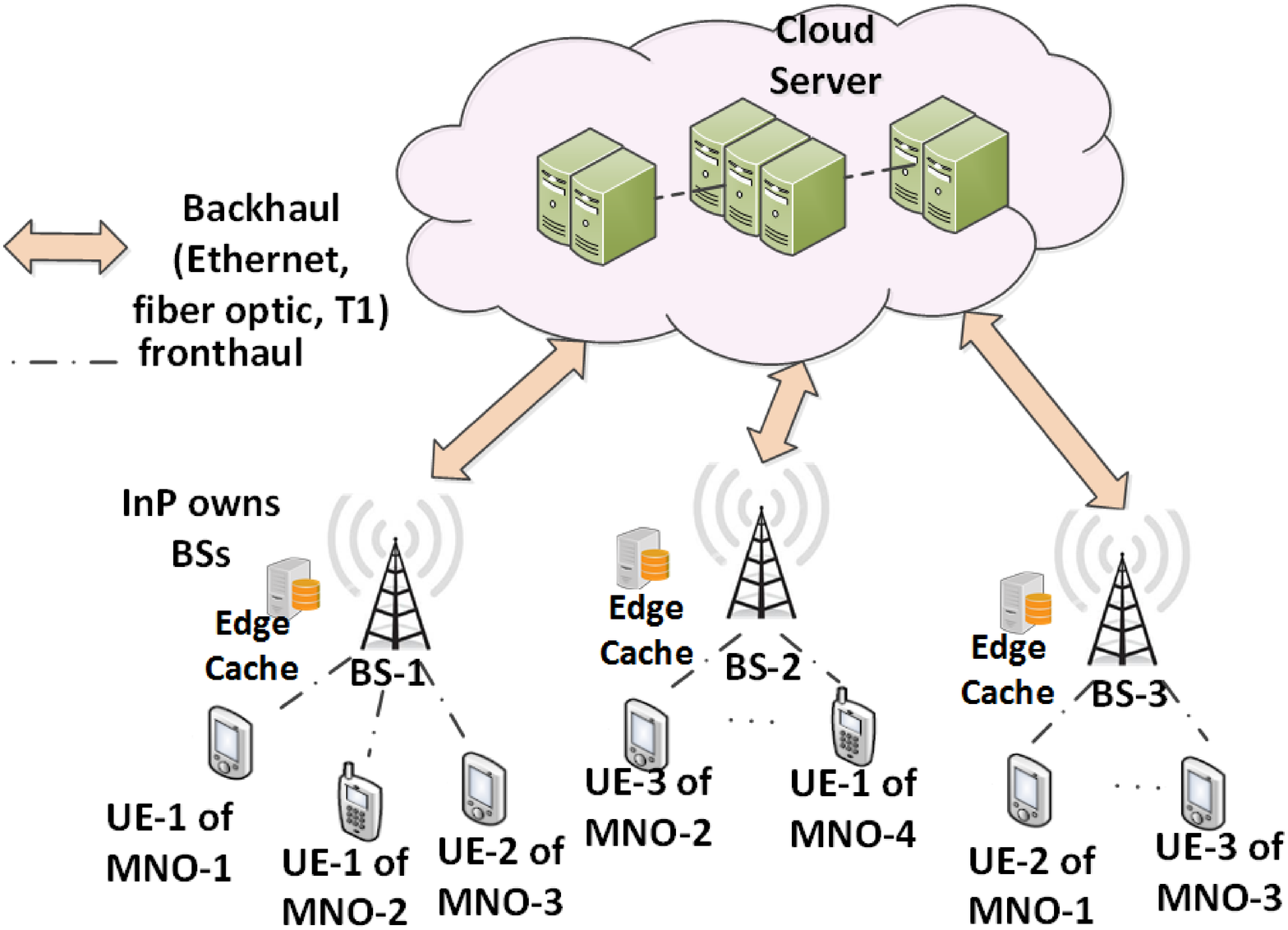}
	\caption{System model.}
	\label{fig:Sys_model}
 \end{center}
 \end{minipage}
\hfill
\begin{minipage}{0.45\textwidth}
\begin{center}
	\includegraphics[width=\textwidth]{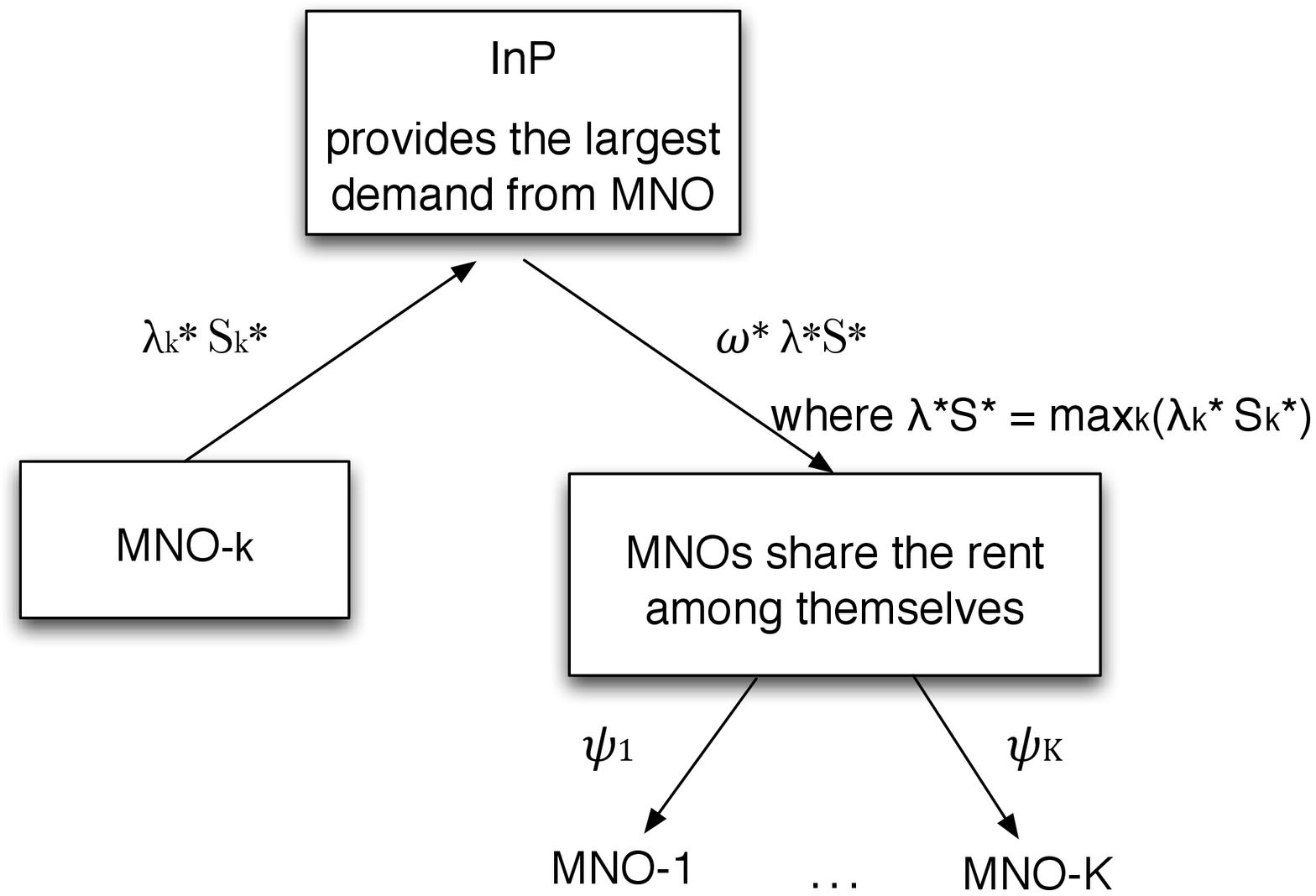}
	\caption{Hierarchical relationship between the MNOs and the InP.}
	\label{fig:Hierarchical_MNO_InP}
 \end{center}
  \end{minipage}
\end{figure}

In this paper, we study (i) the wireless aspects of caching at the BS and (ii) the economic aspects of pricing and the cost sharing of a network infrastructure. We consider MNOs serving users in a common geographic region. The MNOs share a common infrastructure provided by a single InP. The InP deploys cache-enabled BSs and allows multiple MNOs to share both cache storage and BSs. With network virtualization, each physical BS of the InP can be utilized by more than one MNO simultaneously. This means that a single BS can serve UEs subscribed to different MNOs. This is shown schematically in Fig.~\ref{fig:Sys_model}, where each BS can serve UEs of multiple MNOs, i.e., BS-$1$ serves UE of MNO-$1$, MNO-$2$, and MNO-$3$; the BS-$2$ serves UE of MNO-$2$ and MNO-$4$; while the BS-$3$ serves UE of MNO-$1$ and MNO-$3$. The BSs are assumed to connect to the cloud. As such, if a file requested by a UE is not available among its cached files, the BS will fetch the file from the cloud. We assume that each MNO operates over its own spectrum. Thus, there is no inter-operator interference. Each MNO aims to deliver the required quality-of-experience (QoE) to its user, in terms of latency of data delivery, by renting suitable amount of a common infrastructure, regards to the cache size and BS intensity from the InP.

%
%
First, using the results from stochastic geometrical analysis of cellular networks, we model and analyze the delay due to the fronthaul and backhaul of the cellular network while assuming the BSs connect to the network core, i.e., servers in the cloud. Also, we show how the delay can be reduced to a tolerable limit by introducing cache at the BSs and/or by increasing the BS intensity. We formulate a network cost minimization problem through the reduction of the BS intensity and/or the cache storage for each MNO as an optimization problem with probabilistic constraint. Assuming a popularity based caching policy, this optimization problem is transformed into a geometric programming problem which can be solved analytically. 

Second, we model and analyze the economic aspects of pricing and cost sharing in a cache-enabled virtualized cellular network. In particular, we formulate a pricing problem for the InP. The InP aims to maximize its revenue from the MNOs while minimizing the operation cost due to the power consumption at the BSs. The price set by the InP is equivalent to the solution of a one leader - one follower Stackelberg game, where the InP is the leader and the follower is the MNO with the largest demand. The InP caters to the largest number of popular files required by the MNOs to be cached and the solution to the problem is given by the Stackelberg equilibrium. The InP's solution to this game is given by a signomial program, which we transform into a complementary geometric program. Then, we use successive geometric programming approximation, which leads to a simple iterative algorithm, to obtain its optimal solution. Finally, the total rent of an infrastructure needs to be shared by the MNOs. For this purpose, we use the Shapely value to divide the rent fairly among the MNOs.

%
%
%

The relationship between InP and MNOs is illustrated in Fig.~\ref{fig:Hierarchical_MNO_InP}. As we see from the figure that there are two types of games: a Stackelberg game and a cooperative game. In order to compute the Stackelberg equilibrium, we use the backward induction method. Firstly, each MNO computes the optimal strategy in terms of cache intensity ($\lambda^* S^*$)\footnote{We will refer to the product of BS intensity ($\lambda$) and cache size ($S$) as ``\tit{cache intensity}''. This is the number of cached files per unit area of the MNO. It enables us to capture the tradeoff between adding more BSs versus adding more cache space.}. Then, the MNOs ``send'' the best response in terms of its demand for cache intensity to the InP. Secondly, since the infrastructure can be shared among MNOs, the InP will compute the price of the infrastructure so as to maximize its revenue using the highest number of popular files, i.e. cache intensity, required by the MNOs. The InP will then declare the optimal price ($\omega^*$) of infrastructure to all MNOs. Finally, the MNOs will cooperate with each other to share the total rent among each other. At this stage, we introduce the Shapley value as a method to divide the rent fairly among the MNOs ($\psi_k$).

\subsection{Organization}
The rest of the paper is organized as follows. Section~\ref{section:systemmodel} describes the system model of the cache-enabled virtualized cellular network. It also provides the throughput expression for a UE obtained through the stochastic geometric analysis of its downlink signal-to-interference-plus-noise ratio (SINR) coverage probability. The assumptions on the caching policy are also stated and the cache hit probability is analyzed. Section~\ref{subsec:Proactive_Cache} provides the delay modeling for a cache-enabled cellular network considering two types of delays: (i) expected fronthaul delay and (ii) expected backhaul delay. Section~\ref{subsec:MNO-problem-formulation} introduces and solves an optimization model for an MNO (referred to as the {\em MNO problem}) to minimize the {\em cache intensity} subject to the latency constraint at a typical UE. This corresponds to the follower subgame in the Stackelberg game formulation. Section~\ref{subsec:oneInP_oneMNO} presents an optimization problem in order to maximize the revenue of the InP (referred to as the {\em InP problem}, which corresponds to the leader subgame in the Stackelberg game formulation). Section~\ref{subsec:coalition_shapleyvalue} models the cooperation among MNOs for sharing the infrastructure and uses Shapley value to divide the rent among the MNOs. The numerical results are presented in Section~\ref{subsec:results}. The paper is concluded in Section~\ref{subsec:conclusion}.

\section{System Model and Assumptions} \label{section:systemmodel}
\subsection{Network Model}

Consider a single InP and a set $\Kcal$ of MNOs such that $|\Kcal| = K$. The MNOs are assumed to be co-located and serve their UEs in the same area. The MNO wants to rent a fraction of $\Phi_b$ of cache-enabled BSs installed by the InP which are spatially distributed according to a homogeneous Poisson point process (PPP) with spatial intensity $\lambda$ (BSs per unit area). All of the MNOs can simultaneously utilize the common BSs made available by the InP. The intensity of the BSs $\lambda_k$ utilized by the MNO is given by the thinning of $\Phi_b$ of the BSs owned by the InP, which yields another homogeneous PPP $\Phi_k$. We can express the property of thinning and sharing of BSs by $\Phi_k \subseteq \Phi_b$ such that $\cup_k \Phi_k = \Phi_b$ and $\Phi_k \cap \Phi_l$ is not necessarily empty, where $k, l \in \mathcal{K}$ and $k \neq l$. 

The MNOs operate over orthogonal spectrum, thus avoiding inter-operator interference. Every BS and UE are assumed to be equipped with a single antenna. Each BS is assumed to transmit with power $p$. For an MNO, the BSs serve the corresponding UEs over its own spectrum in a time-sharing manner. Thus, the BS serves one UE in a given time slot for the assigned spectrum. A UE subscribed to an MNO associates with the nearest available BS that the MNO has rented from the InP. However, we do not consider explicitly the BSs that the MNO has rented from the InP. Instead, we assume that the MNO is free to use any fraction of the total available BSs. The net intensity of the BSs that a typical UE of the MNO can associate itself with is $\lambda_A = \lambda_k$. The UEs subscribed to MNO-$k$, the set of which is denoted by $\Phi_{u_k}$, are assumed to be spatially distributed according to a homogeneous PPP with spatial intensity $\xi_k$. Let $\eta_k \in (0,1)$ be the activity level of a UE. The two point processes $\Phi_k$ and $\Phi_{u_k}$ are assumed to be independent. We assume that each MNO has a bandwidth of $W_k$ Hz, which is divided into $L_k$ subchannels. Each BS operates in one of the $L_k$ available subchannels randomly assigned to it by the MNO. Thus, the intensity of interfering BSs is given by $\lambda_I = \frac{\lambda_k}{L_k}$.

\subsection{Downlink SINR Coverage Probability and Throughput}
\label{sec:stogeoana}
Consider a single MNO-$k$. To avoid the notational clutter, without any cause for ambiguity, we will drop the subscript $k$ from all the MNO parameters in this part and the rest of the sections until Section~\ref{subsec:MNO-problem-formulation} when we explicitly consider multiple MNOs. Until then, we are exclusively dealing with a single MNO. Without loss of generality, we consider a typical UE of MNO to be  located at the origin, which associates with the nearest BS. For convenience, let us label the nearest BS as BS-$0$. We assume that the message signal undergoes Rayleigh fading with the channel gain given by $g_0$. Furthermore, let $\alpha > 2$ denote the path-loss exponent for the path-loss model $r_0^{-\alpha}$, where $r_0$ is the distance between the typical UE and the nearest BS-$0$, $0 \in \Phi_b$. Finally, let $\sigma^2$ denote the noise variance and $p$ denote the transmit power of all the BSs of the MNO. The downlink ${\rm SINR}$ at the typical UE is:
${\rm SINR} = \frac{g_0 r_0^{-\alpha}p}{I + \sigma^2}.$ The interference experienced by a typical UE associated with BS-$0$ comes from the transmit signal from other BSs to the UEs in the same time slot. Thus, $I = \sum_{j\in\Phi_b \backslash \{0\}} g_j r_j^{-\alpha}p$. Here $g_j$ is the channel gain between the typical UE and interfering BS-$j$, and $r_j$ is the distance between the typical UE and the interfering BS-$j$, where $j\in\Phi_b \backslash \{0\}$. 

For a given threshold $\bar{T}$, the SINR coverage probability for the typical UE is defined as $P_c = \mathrm{Pr}({\rm SINR} > \bar{T})$. Following the approach given in \cite[Theorem 1]{Andrews2011}, the analytical expression for $P_c$ under our system assumptions is given by \cite[Prop. 1]{Tachporn_TMC2017} as  $P_c = \pi \lambda_A \int_0^{\infty}\exp \{-(\bar{A} z + \bar{B} z^{\alpha /2})\} dz,$ where the coefficients $\bar{A}$ and $\bar{B}$ are given by $\bar{A} = \pi [\lambda_I(\beta -1) + \lambda_A]$ and $\bar{B} = \frac{\bar{T} \sigma^2}{p}$. Here, $\lambda_A = \lambda$,  $\lambda_I = \lambda/L$, and $\beta = \frac{2 (\bar{T}/p)^{2/\alpha}}{\alpha} \mathbb{E}_{g}[g^{2/\alpha} (\Gamma(-2/\alpha, \bar{T} g/p))- \Gamma(-2/ \alpha)]$. Using a simple closed-form approximation given in \cite[Eqn. 4]{Sudarshan2016}, we can evaluate $P_c$ as 
\begin{align}
P_c \simeq  \; \pi \lambda_A \left[ \bar{A}+ \frac{\alpha}{2} \frac{\bar{B}^{2/\alpha}}{\Gamma\big(\frac{2}{\alpha}\big)} \right]^{-1} 
= \left[ 1 + \frac{\beta - 1}{L} + \frac{\alpha}{2 \pi \lambda \Gamma\big(\frac{2}{\alpha}\big)} \left( \frac{\bar{T}\sigma^2}{p} \right)^{2/\alpha} \right]^{-1},
\label{eqn:coverage-approx}
\end{align}
where $\Gamma(z)$ is Gamma function. 
For interference-limited case, when $\sigma^2 \to 0$ or when $\lambda \to \infty$, (\ref{eqn:coverage-approx}) simplifies to 
\beq
	P_c \simeq \frac{L}{\beta + L - 1}.
\label{eqn:coverage-approx-interferencelimited}
\eeq
Equation (\ref{eqn:coverage-approx-interferencelimited}) is independent of $\lambda$. Also, as $L \to \infty$, we have from (\ref{eqn:coverage-approx-interferencelimited}) that $P_c \to 1$. Next, we define the {\em throughput} of a typical UE served by the nearest BS as
\beq
G = \frac{P_c W}{L} \log_2 (1+\bar{T}), 
\label{eqn:goodput}
\eeq
where $P_c$ is the downlink coverage probability, and $W$ is bandwidth. We can approximate the throughput in (\ref{eqn:goodput}) by using (\ref{eqn:coverage-approx}) in general, or (\ref{eqn:coverage-approx-interferencelimited}) for the interference-limited case. We see that increasing $L$ improves the coverage probability but reduces the throughput.

\subsection{Caching Policy}
In order to enable edge caching, let each BS have $S$ amount of cache. That is, each BS can store $S$ files in its storage. For simplicity, we assume that all the files are of equal size. Let $\mathcal{F} = \{f_1 \ldots, f_{F}\}$ be the set of files available for caching, where $F = |\mathcal{F}|$ is the total number of files. Based on the file popularity distribution and cache replacement policy at the edge, let $\Scal \subseteq \Fcal$ be the set of files cached at each BS, where $|\Scal| = S$ is the cache size. If a random file $f\in\Fcal$ is requested by a UE, then let the probability that the file-$f$ is available at the BS cache be $P_{\text{hit}}(S) = \Pr(f \in \Scal)$, which we refer to as ``\tit{hit probability}.'' 

We assume that the cache policy is to store the $S$ most popular files from $\Fcal$. We can model the popularity of the files by Zipf distribution given by, $p_{d} = \frac{1/d^{\nu}}{\sum_{j=1}^{F} 1/j^{\nu}},$ where $p_d$ is the probability of $d$-th most popular file being requested and the exponent $\nu>0$ reflects the skewness of the content popularity distribution. The larger of value $\nu$, the fewer popular contents hold a majority of the content requests. The probability that the requested file $f \in \Fcal$ is stored in the cache is $P_{\text{hit}}(S) = \Pr(d \leq S)$, where $d$ is the random popularity rank of the file $f$. We observe that $\Pr(d \leq S)$ is the cumulative distribution function (CDF) of the Zipf distribution. Hence, we can express $P_{\text{hit}}(S)$ as 
 \beq
 P_{\text{hit}}(S) = \frac{\sum_{d=1}^S 1/d^{\nu}}{\sum_{j=1}^{F} 1/j^{\nu}} = \frac{H_{S,\nu}}{H_{F,\nu}}.
 \label{eqn:hit-probability}
 \eeq
In (\ref{eqn:hit-probability}), we have concisely expressed $P_{\text{hit}}$ using generalized harmonic numbers, $H_{S,\nu}$ and $H_{F,\nu}$, where
\beq 
H_{S,\nu} = \sum_{n=0}^{S-1} \frac{1}{(n+1)^\nu}.
\label{eqn:def-genearlized-harmonic-sum}
\eeq
The $H_{F,\nu}$ is defined similarly. 

\subsection{Asymptotic Approximation for the Cache Hit Probability}

In order to facilitate subsequent analysis, we now give the following lemma on the asymptotic approximation\footnote{Here $f(x) \sim g(x)$ if and only if $\lim_{x\to \infty}\frac{f(x)}{g(x)} = 1$.} for the hit probability: 
 
\begin{lemma}
When the size of cache $S$ is large and $\nu \neq 1$, the probability that the requested file $f\in \Fcal$ is in the cache is asymptotically given by
\beq
P_{\text{hit}}(S) \sim \frac{1}{H_{F,\nu}} \left[ \zeta (\nu) - \frac{(S+1)^{1-\nu}}{\nu -1}\right],
\label{eqn:Phit_asymptotic}
\eeq
where $\zeta(\nu)$ is the Riemann zeta function. 
\label{lemma:hit-prob}
\end{lemma}

\begin{IEEEproof}
See \tbf{Appendix}.
\end{IEEEproof}

Here (\ref{eqn:Phit_asymptotic}) is asymptotic in the sense that larger values of $S$ result in a greater accuracy of the formula. Note that although it was stipulated that $\Re(\nu) > 1$ during the definition of the Hurwitz zeta function in (\ref{eqn:hurwitz-zeta-fun}), the Riemann zeta function $\zeta(\nu)$ has a unique analytic continuation to the entire complex plane, excluding $\nu = 1$, which corresponds to a simple pole~\cite{Sondow2017a}. Similar analytical continuation holds for Hurwitz zeta function as well~\cite{Sondow2017b}. Thus, so long as $\nu \neq 1$, the formula in (\ref{eqn:Phit_asymptotic}) is applicable for any Zipf's exponent $\nu > 0$. 

\begin{figure}[h]
\begin{minipage}{0.45\textwidth}
\begin{center}
	\includegraphics[width=\textwidth]{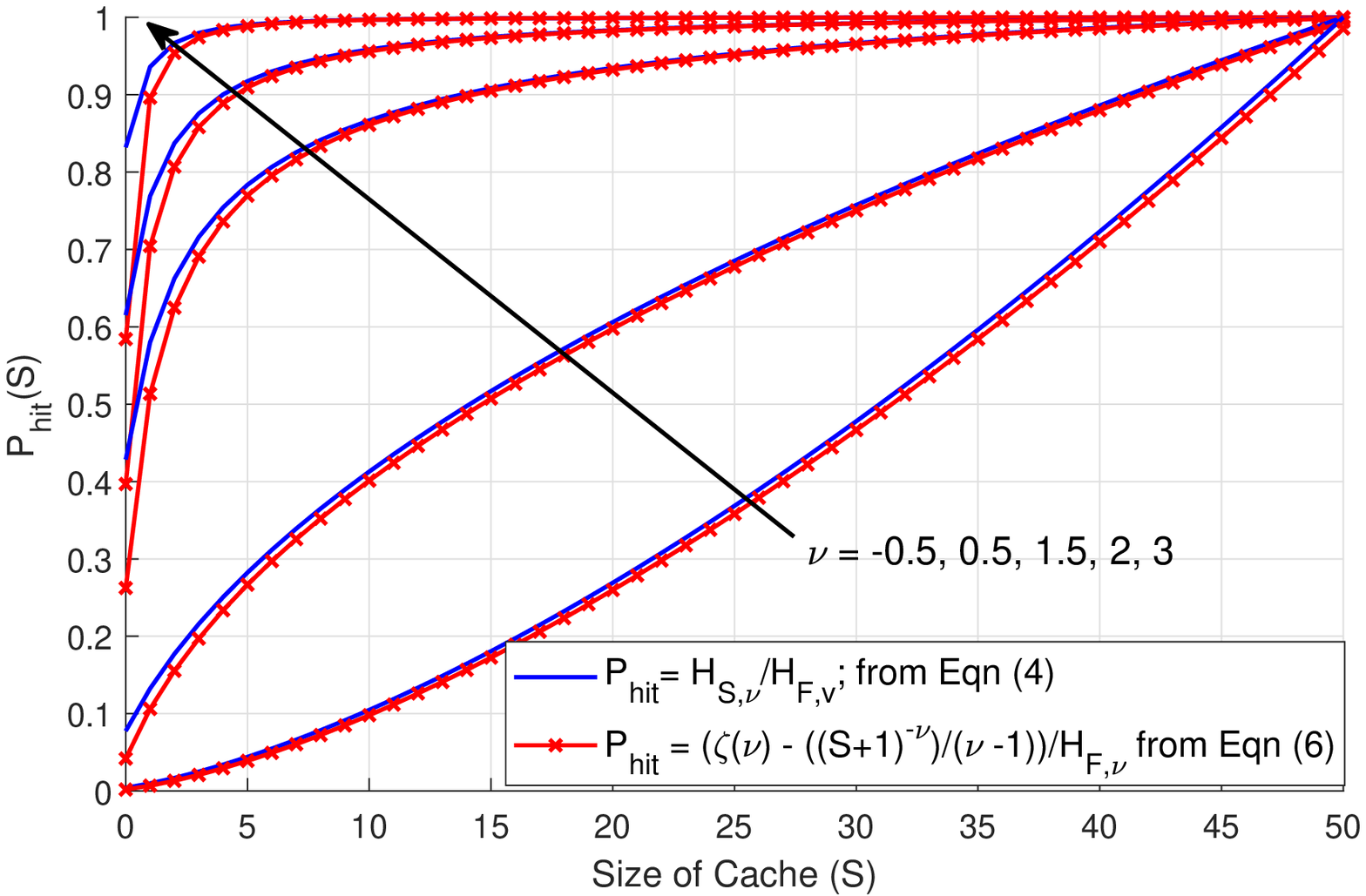}
	\caption{Hit probability versus size of cache ($S$).}
	\label{fig:Hit_prob:S}
 \end{center}
 \end{minipage}
\hfill
\begin{minipage}{0.45\textwidth}
\begin{center}
	\includegraphics[width=\textwidth]{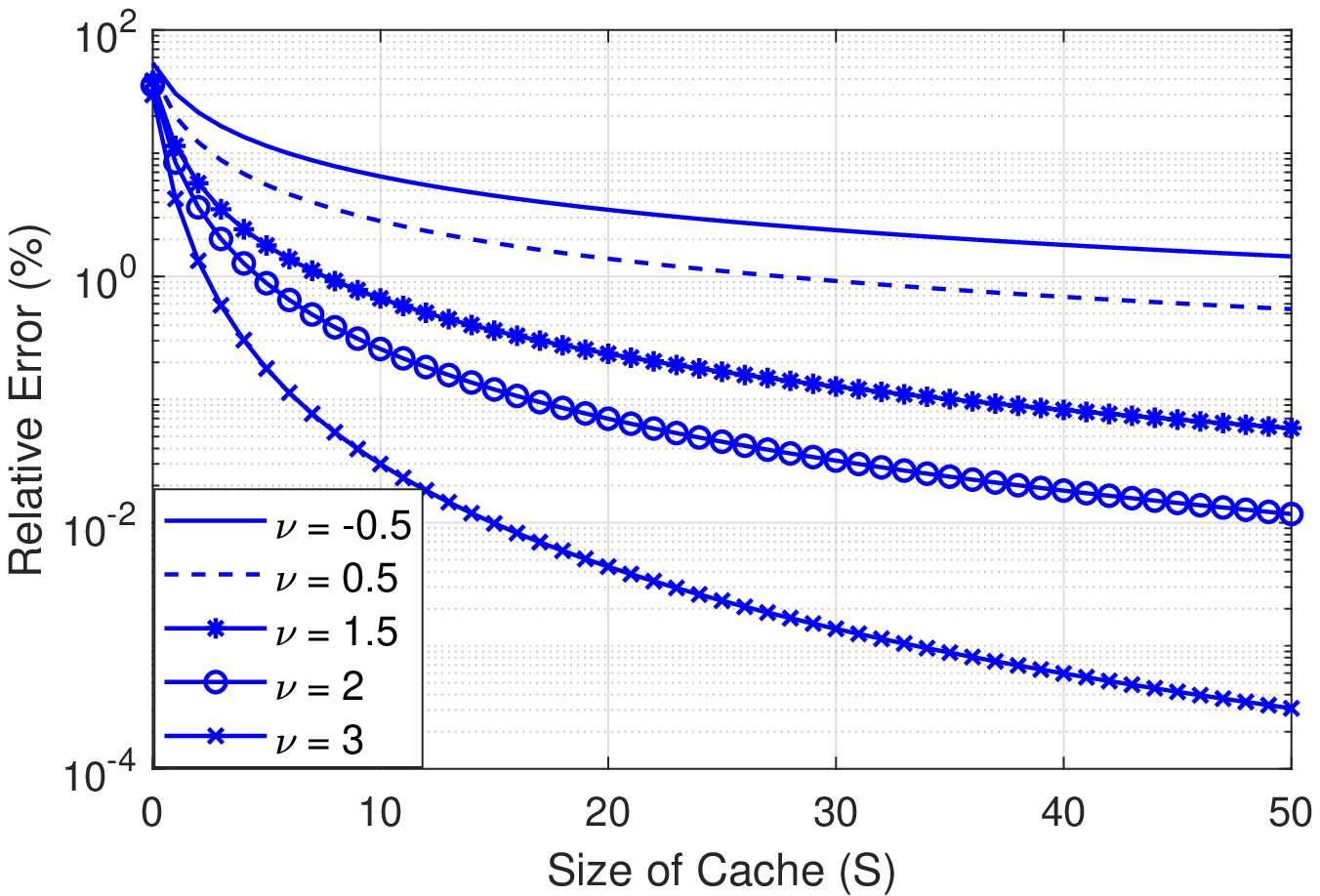}
	\caption{Relative error versus size of cache ($S$).}
	\label{fig:Error:S}
 \end{center}
  \end{minipage}
\end{figure}

In reality, Zipf's exponent is found to be close to but never exactly equal to $1$. There is  no consensus on the actual setting of $\nu$ value~\cite{Cha2009,Fricker2012}, with the considered value varying widely between $\nu \in [0.5, 2.5]$. Also, we expect the size of the cache to be $1<<S<<F$, so the above approximation holds with very small margin of error. In Fig.~\ref{fig:Hit_prob:S}, we compare the hit probability versus size of cache using the exact values of $H_{s,\nu}$ as obtained from (\ref{eqn:hurwitz-zeta-harmonic-sum-2}) and the asymptotic approximation of $H_{s,\nu}$ from (\ref{eqn:harmonic-sum-asymptotic}) when $F = 10^3$. The relative error is shown in Fig.~\ref{fig:Error:S}. We observe that the relative error decreases with increasing cache size. The error tends to decrease more rapidly for larger values of $\nu$. For $\nu \geq 0.5$, the relative error is less than $1 \%$ when $S \geq 30$, while for $\nu \geq 1.5$ the relative error is less than $1 \%$ when $S \geq  10$. Finally, we also note that the formula is applicable even when $\nu < 0$.


\tbf{Remark 1.} Applying the equation (\ref{eqn:harmonic-sum-asymptotic}) to $H_{F,\nu}$, we see that
\beq 
P_{\text{hit}}(S) \sim \frac{ \zeta (\nu) - \frac{(S+1)^{1-\nu}}{\nu -1}} {\zeta (\nu) - \frac{(F+1)^{1-\nu}}{\nu -1}}. 
\label{eqn:Phit_asymptotic1}
\eeq
Thus, for a fixed value of $S$, $P_{\text{hit}}$ decreases with increasing $F$. 

\tbf{Remark 2.} Let $s = \frac{S+1}{F+1}$ be the fixed fraction of files cached at the BS. Dividing the numerator and denominator of (\ref{eqn:Phit_asymptotic1}) by $(F+1)^{1-\nu}/(\nu -1)$, we see that as $F \to \infty$ and as $S$ changes such that the fraction $s$ is fixed, we obtain
\beq
P_{\text{hit}} \sim s^{1-\nu},
\eeq
when $0<\nu<1$. This result has an important implication in that, to obtain a desired hit probability, the fraction of total files that need to be cached at the BS is 
\beq
s \sim P_{\text{hit}}^{\frac{1}{1-\nu}},
\label{s:P_hit}
\eeq
so long as $0 < \nu < 1$. If $\nu > 1$, then $P_{\text{hit}} \sim 1$ as $F \rightarrow \infty$. 

\section{Delay Modeling in a Cache-Enabled Downlink Cellular Network}
\label{subsec:Proactive_Cache}


With caching at the BSs, for cellular downlink communications, there are two types of delays: fronthaul delay and backhaul delay. In the following we will characterize the expected value of these two delays. We will see that in a cache-enabled cellular system, the fronthaul presents the ultimate bottleneck, which can only be remedied by deploying more BSs.

\subsection{Expected Fronthaul Delay}
\label{subsec:fronthaul_delay}

The delay in the transmission of a file between BS and UE is referred to as the \tit{fronthaul delay}. If a file requested by a UE is available in the cache of the serving BS, then the delay incurred during transmission of the file is only due to the fronthaul. This delay is contributed by a number of factors, including finite channel capacity, transmission success rate, size of the file, and the number of UEs in a cell.
The potential delay due to the channel is the reciprocal of the throughput, $1/G$, in seconds per bit. If there are $N$ UEs in the cell being served simultaneously, since the BS serves the users in a time-sharing manner, the throughput per UE is $G/N$. Thus, the potential delay for a UE is $N/G$. Hence, in order to transfer a file of fixed size $x_f$, the total fronthaul delay, is $D_{\text{fh}} = \frac{N x_f}{G}$. Here, only $N$ is the random variable. The expected number of UEs inside an average Voronoi cell formed by the PPP of the BS, $\Phi_b$, is given by $\frac{\xi}{\lambda}$~\cite[Eqn 20]{Arvanitakis2015}. Thus, the number of UEs being served is $\Ebb[N] = \frac{\eta \xi}{\lambda}$,
where $\xi$ is the spatial intensity of the UEs and $\eta \in (0,1)$ is the probability that a UE will request service from the BS. When $\eta$ is small, the MNO can retain a large number of  UEs to be potentially served by a single BS by its limited infrastructure. Hence, the average fronthaul delay is given by
\beq 
\Ebb[D_{\text{fh}}] = \frac{\Ebb[N] x_f}{G} = \frac{\eta \xi x_f}{\lambda G}.
\label{eqn:fronthaul-delay}
\eeq 

Equation (\ref{eqn:fronthaul-delay}) confirms our intuitive understanding that BS densification leads to lower fronthaul delay. Since we are using the throughput as defined by (\ref{eqn:goodput}), equation (\ref{eqn:fronthaul-delay}) represents the average worst-case delay for the user. Optimizing the system parameters with respect to this worst-case ensures that the user experiences lower delay in practice.

\subsection{Expected Backhaul Delay}
\label{subsec:backhaul_delay}
When the requested file from the UE is \tit{not} available in the cache of the serving BS, then the file needs to be fetched from cloud server to the BS. The delay in the transmission of a file between the BS and the cloud server is referred to as the \tit{backhaul delay}. Let us assume that the BSs connect to cloud via a backhaul link. Assuming that the application provider has $m$ servers, we can model the process of a BS fetching the contents from the cloud server as a $G/G/m$ queue. Let $\tau$ be the mean service time at a single server. That is, $\tau$ is the average time taken for the server to deliver $x_f$ bits of information to the BS. Then, the expected backhaul delay is given by the total expected sojourn time of the $G/G/m$ queue as~\cite[Eqn 2.14]{Whitt1993}
\beq 
\Ebb[D_{\text{bh}}] \approx \left( \frac{c_a^2 + c_s^2}{2} \right) \Ebb[\widehat{W}(M/M/m)] + \tau,
\label{eqn:backhaul-delay}
\eeq 
where $\Ebb[\widehat{W}(M/M/m)] \approx \tau (\rho^{\sqrt{2(m+1)}-1})/(m(1-\rho))$ is the expected waiting time in an $M/M/m$ queue. Here, $\phi$ denotes the mean arrival rate of file requests to the server, $\mu = m/\tau$ is service rate of the server, and $\rho = \phi/(m \mu)$ is the server utilization. Also, $c_a$ and $c_s$ are coefficients of variations of the inter-arrival time and the service time, respectively. For the stability of the cloud server queue, the condition $\rho < 1$ must be satisfied. Note that we have implicitly assumed that the caching at BS by an MNO does not change the request arrival rate, $\phi$, at the server. Also, when $m=1$, the above approximation yields $\Ebb[D_{\text{bh}}] \approx \left( \frac{c_a^2 + c_s^2}{2} \right) \left( \frac{\rho}{1-\rho} \right) \tau + \tau$.

\subsection{Expected Total Delay}
\label{subsec:Total_delay}
 The delay experienced by a UE while downloading a file is only due to the fronthaul, $D_{\text{fh}}$, if the requested file is already cached at its serving BS. If this is not the case, then the delay experienced by the UE is the sum of the fronthaul and backhaul delay, $D_{\text{fh}} + D_{\text{bh}}$. Since the availability of a file in the BS cache is given by the hit probability, $P_{\text{hit}}$, the expected total delay is given by the law of total expectation as
 \begin{align}
\Ebb[D] &= \Ebb[D_{\text{fh}}] \, P_{\text{hit}} + \Ebb[D_{\text{fh}} + D_{\text{bh}}] \, (1 - P_{\text{hit}}) = \Ebb[D_{\text{fh}}] + \Ebb[D_{\text{bh}}] (1 - P_{\text{hit}}). \label{eqn:expected-total-delay}
 \end{align}
 
 We have modeled the expected fronthaul and backhaul delay as given by (\ref{eqn:fronthaul-delay}) and (\ref{eqn:backhaul-delay}), respectively. Since $0 \leq P_{\text{hit}} \leq 1$, notice that the average total delay is bounded by $\Ebb[D_{\text{fh}}] \leq \Ebb[D] \leq \Ebb[D_{\text{fh}}] + \Ebb[D_{\text{bh}}]$. Since $P_{\text{hit}}$ depends on the cache size $S$, this implies that the minimum expected delay that we can achieve by changing only the cache size is $\Ebb[D_{\text{fh}}]$. This bound also suggests that we cannot impose an arbitrarily lower bound on the total delay by only adding cache to the BSs, which allows us to eliminate the backhaul delay only. In terms of total delay, the fronthaul presents the ultimate bottleneck for cache-enabled cellular systems; therefore, we also need to increase the number of BSs deployed to lower the fronthaul delay.

\section{Optimization of Cache Intensity: The MNO Problem}
\label{subsec:MNO-problem-formulation}
In this section, we will deal with the minimization of cache intensity for an MNO. We will transform the problem into a geometric programming problem and give an exact solution.

\subsection{Optimization Problem Formulation}

We formulate an optimization problem so as to minimize the cost of deploying the amount of cache per unit area (cache intensity) while satisfying the delay constraint for a typical UE. The optimization problem for each MNO is as follows:
\begin{align}
\label{eqn:min_lambda_S}
(P0) \quad \text{min}_{\lambda, S} & \quad \omega \lambda S \\
\text{s.t}. & \quad \text{Pr}(D \geq D_{\text{th}}) \leq \gamma, 
\label{eqn:geometric_prog} \\
& \quad S \leq F, \label{eqn:cache-constraint}
\end{align}
where $\lambda \geq 0$ and $S \geq 0$. Here, (\ref{eqn:geometric_prog}) is a probabilistic constraint that limits the latency above some threshold value $D_{\text{th}}$ to probability $\gamma \in (0,1)$. The $\gamma$ is set to be some small value. The $\omega$ is the price per unit cache intensity as set by the InP. To make the problem more tractable, we have from Markov's inequality 
\begin{align}
\text{Pr}(D \geq D_{\text{th}}) \leq \frac{\Ebb[D]}{ D_{\text{th}}}.
\label{eqn:Markov_in_Equal}
\end{align}

Using the Markov's inequality in (\ref{eqn:Markov_in_Equal}), we can substitute the probabilistic constraint in (\ref{eqn:geometric_prog}) by another constraint 
\beq 
\Ebb[D] \leq \gamma D_{\text{th}}.
\label{eqn:tot_cache_Th}
\eeq 
If the inequality in (\ref{eqn:tot_cache_Th}) is satisfied, then it implies that the inequality in (\ref{eqn:geometric_prog}) is also satisfied. Substituting the expression for $\Ebb[D]$ from (\ref{eqn:expected-total-delay}) into (\ref{eqn:tot_cache_Th}),  after some algebra  we obtain
\begin{align}
 1 - \frac{\gamma D_{\text{th}} - \Ebb[D_{\text{fh}}] }{\Ebb[D_{\text{bh}}]} &\leq P_{\text{hit}}(S). 
\label{eqn:expected_cache2}
\end{align}

Here, in (\ref{eqn:expected_cache2}), we have succeeded in modifying the statement about the delay constraint into an equivalent statement concerning the cache size. Since $P_{\text{hit}}(S) \leq 1$, for the sake of consistency, in (\ref{eqn:expected_cache2}) it must be the case that
\beq 
\mathbb{E}[D_{\text{fh}}] \leq \gamma D_{\text{th}},
\label{eqn:sufficient-condition}
\eeq
so that the left-hand-side of (\ref{eqn:expected_cache2}) is less than the unity. Thus we have a fundamental lemma about cache-enabled cellular system as stated in the next subsection.

\subsection{Trade-offs Between Cache Storage and BS Intensity}
 
 \begin{lemma}
The constraint in (\ref{eqn:geometric_prog}) is feasible for some $S$ such that $S \leq F$ if and only if $\mathbb{E}[D_{\text{fh}}] \leq \gamma D_{\text{th}}$.
 \label{lemma:feasibility-of-delay-constraint}
 \end{lemma}
 
The \tbf{Lemma~\ref{lemma:feasibility-of-delay-constraint}} is fundamental in the sense that it must be satisfied regardless of how we model the fronthaul delay or the hit probability or the caching policy, since its veracity does not depend on such specific modeling assumptions. \tbf{Lemma~\ref{lemma:feasibility-of-delay-constraint}} gives us the necessary and sufficient condition under which both the constraints (\ref{eqn:geometric_prog}) and (\ref{eqn:cache-constraint}) can be feasible. Although the delay constraint in (\ref{eqn:tot_cache_Th}) is over the total delay, we see that the fronthaul delay plays the most crucial part. Equation (\ref{eqn:sufficient-condition}) also implies that we cannot set $D_{\text{th}}$ arbitrarily low for a given value of $\gamma$. Since $\mathbb{E}[D_{\text{fh}}]$ is a constant for given $\lambda$, we see that $\gamma$ and $D_{\text{th}}$ should be at least inversely proportional to each other. If we have $\mathbb{E}[D_{\text{fh}}] = \gamma D_{\text{th}}$, then the cache size is $S = F$. That is, the BS should cache all the available files in $\Fcal$. This is certainly unrealistic in practice. Hence, realistically, it should be the case that $\mathbb{E}[D_{\text{fh}}] < \gamma D_{\text{th}}$ so that $S < F$. If we substitute the expression for $\mathbb{E}[D_{\text{fh}}]$ from (\ref{eqn:fronthaul-delay}) in (\ref{eqn:sufficient-condition}), we obtain a minimum bound for the BS intensity required for the feasibility of (\ref{eqn:sufficient-condition}), and hence (\ref{eqn:geometric_prog}), as
\beq
\lambda \geq \frac{\eta \xi x_f}{\gamma D_{\text{th}} G}.
\label{eqn:lambda-lower-bound}
\eeq
This gives us a relationship between the intensity of the BSs $\lambda$ and the intensity of the UEs $\xi$ for the delay constraint to be feasible for some $S$ such that $S<F$. If this condition is violated, then the required $S$ will be greater than $F$, leading to a contradiction. We can also express this bound in terms of the average number of UE served per BS, $\frac{\xi}{\lambda}$. This leads to our first proposition:
 
 \begin{proposition}
 For the delay constraint in (\ref{eqn:sufficient-condition}) to be satisfied for some $S$ such that $S \leq F$, the average number of UE per BS, $\frac{\xi}{\lambda}$, must be lower than $\frac{\gamma D_{\text{th}} G}{\eta x_f}$. That is, we have the upper bound $\frac{\xi}{\lambda} \leq \frac{\gamma D_{\text{th}} G}{\eta x_f}$.
 \label{prop:lambda-lower-bound}
 \end{proposition}
 
In practice, we expect $S << F$. Therefore, the above proposition gives us a simple condition under which caching at the BS makes physical sense in order to meet the required delay constraint. As with \tbf{Lemma~\ref{lemma:feasibility-of-delay-constraint}}, the validity of \tbf{Proposition~\ref{prop:lambda-lower-bound}} does not depend on the caching policy. Indeed, if $\lambda$ is held fixed, implying that we cannot change the fronthaul delay and can only change the total delay by changing the cache size, then we can solve our problem (P0) for popularity based caching when $\nu < 1$ by substituting (\ref{s:P_hit}) into  (\ref{eqn:expected_cache2}). This strategy is not possible when $\nu > 1$. 
Hence, we need to vary both $\lambda$ as well as $S$. This results in problem (P1) given in the next subsection.

\subsection{Optimal Strategy of MNO}
\label{sec:best-response-MNO}
We will now solve the general problem when both $\lambda$ and $S$ are jointly optimized. Given the popularity-based caching, for large $S$, and $S < < F$, we can recognize the optimization problem (\ref{eqn:min_lambda_S})-(\ref{eqn:geometric_prog}) as a geometric programming problem~\cite{Duffin1967}. In the following, we will first express the primal problem in (\ref{eqn:geometric_prog}) in the standard form of a geometric program, after which, we will give the solution to the problem via its dual problem.

First, we expand $\mathbb{E}[D]$ in (\ref{eqn:expected-total-delay}) using (\ref{eqn:fronthaul-delay}) and (\ref{eqn:Phit_asymptotic}) in terms of $\lambda$ and $S$ as 
\begin{align}
 \Ebb[D] &=  \frac{\eta \xi x_f}{G \lambda} + \Ebb[D_{\text{bh}}] \left[ 1-\frac{1}{H_{F,\nu}} \left( \zeta (\nu) - \frac{(S+1)^{1-\nu}}{\nu -1}\right) \right] = C_1 + \frac{C_2}{\lambda} + C_3 (S+1)^{1-\nu}, 
\label{eqn:totcache_withC}                         
\end{align}
where $C_1 = \Ebb[D_{\text{bh}}]\left( 1 - \frac{\zeta(\nu)}{H_{F,\nu}}\right)$, $C_2 = \frac{\eta \xi x_f}{G}$, and $C_3 = \frac{ \mathbb{E}[D_{\text{bh}}]}{(\nu -1) H_{F,\nu}}$. Substituting (\ref{eqn:totcache_withC}) in the constraint (\ref{eqn:tot_cache_Th}) and observing that $S + 1 \approx S$ for large $S$, we obtain, $C_1 + \frac{C_2}{\lambda} + C_3 S^{1-\nu} \leq \gamma D_{\text{th}}   \Longleftrightarrow \left( \frac{C_2}{\gamma D_{\text{th}} - C_1} \right) \frac{1}{\lambda} + \left(\frac{C_3}{\gamma D_{\text{th}} - C_1}\right)S^{1-\nu} \leq 1$. Therefore,
\begin{align}
& A \lambda^{-1} + V S^{1-\nu} \leq 1,
\label{eqn:geometric_constraint}
\end{align}
where $A = \frac{C_2}{\gamma D_{\text{th}} - C_1}$ and $V= \frac{C_3}{\gamma D_{\text{th}} - C_1}$. Now, substituting (\ref{eqn:geometric_constraint}) into the constraint in (\ref{eqn:tot_cache_Th}), and substituting (\ref{eqn:lambda-lower-bound}) in the constraint (\ref{eqn:cache-constraint}), we can express the primal problem (\ref{eqn:geometric_prog}) as a geometric program:

\begin{proposition}
\label{prop:P1_transformgeometric}
Let the caching policy be based on the popularity of files. Assuming $S$ to be sufficiently large and that the constants $A > 0$, $V > 0$, and $\nu \neq 1$, we can transform the problem (P0) into a geometric programming problem 
\begin{align}
\label{eqn:min_lambda_t}
(P1) \quad \text{min}_{\lambda, S} & \quad g = \omega \lambda S \\
\text{s.t}. & \quad A \lambda^{-1} + V S^{1-\nu} \leq 1, \\
& \quad R \lambda^{-1} \leq 1,
\end{align}
where $R = \frac{C_2}{\gamma D_{\text{th}}}$.
\end{proposition}

The optimization problem in (P1) can be solved analytically. In geometric programming, when the orthogonality and normality conditions with dual variables $\delta_i$ are satisfied, the maximum of dual function is equal to the minimum of primal function $g$~\cite{Duffin1967}. As such, we can express the dual maximization problem as

\begin{align}
\label{eqn:dual_function}
\text{max}_{\boldsymbol{\delta}} & \: q = \bigg(\frac{\omega}{\delta_1}\bigg)^{\delta_1} \bigg(\frac{A}{\delta_2}\bigg)^{\delta_2} \bigg(\frac{V}{\delta_3}\bigg)^{\delta_3} \bigg(\frac{R}{\delta_4}\bigg)^{\delta_4} (\delta_2 + \delta_3)^{\delta_2 + \delta_3} (\delta_4)^{\delta_4},\\ 
\text{s.t}. &\quad \delta_1 = 1, \label{eqn:geo_normality} \\
        & \begin{pmatrix}
         1 & -1 & 0 & -1\\
         1 & 0 & 1 - \nu & 0
        \end{pmatrix}   
        \begin{pmatrix}
        \delta_1 \\ \delta_2 \\ \delta_3 \\ \delta_4
		  \end{pmatrix}       
      = 0,
\label{eqn:geo_orthogonality}
\end{align} 
where $\delta_i \geq 0$ for $i = 1, \ldots, 4$. The degree of difficulty of this geometric program is 1. In our case, (\ref{eqn:geo_normality}) gives the normality condition while (\ref{eqn:geo_orthogonality}) gives the orthogonality condition. In geometric programming, we focus on finding the optimal point of the dual variables $\boldsymbol{\delta}^{*} = (\delta_1^{*}, \delta_2^{*}, \delta_3^{*}, \delta_4^*)$ that maximizes the dual function $q$ subject to the orthogonality and normality conditions. Note that this dual problem is a convex program with a concave objective function and linear constraints.

Using (\ref{eqn:geo_normality}) and (\ref{eqn:geo_orthogonality}), we can directly solve for $\boldsymbol{\delta}^{*}$. Here, matrix multiplication from (\ref{eqn:geo_orthogonality}) yields 
\begin{align*}
\delta_1 - \delta_2 - \delta_4 = 0, \quad \mathrm{and}\; \quad \delta_1 + (1-\nu)\delta_3 =0.
\end{align*}

Since $\delta_1 = 1$, we have $\delta_3 = \frac{1}{\nu - 1}$ and $\delta_2 + \delta_4 = 1$. Let $\delta_2 = r$, so that $\delta_4 = 1- r$. Since $\delta_2 \geq 0$ and $\delta_4 \geq 0$, we then have a bound over $r$ as $0 \leq r \leq 1$. Substituting the values of $\delta$'s in the dual problem, we obtain a simpler problem constrained over a single variable $r$ as given in (\ref{eqn:modified-dual-problem})-(\ref{eqn:bound-on-r}). To find the optimal $r$, we first take the logarithm of $q$ in (\ref{eqn:modified-dual-problem}) and differentiate with respect to $r$. Since $A, R, \nu$ are all positive, we obtain
\beq 
\frac{\ud \log(q)}{\ud r} = \log\left(\frac{A}{rR}\right) + \log\left(\frac{1+r(\nu - 1)}{\nu - 1}\right). 
\label{eqn:derivative-of-log-q}
\eeq
Solving the optimality condition $\frac{\ud \log(q)}{\ud r} = 0$ for $r$, we obtain the maxima at $r = \frac{1}{(\nu - 1) \left( \frac{R}{A} -1 \right)}$.
However, since $r$ is bounded between $0 \leq r \leq 1$, we have the optima of the modified dual problem at
\beq 
r^* = \max \left(0, \min \left(1, \left[(\nu - 1) \left( \frac{R}{A} -1 \right) \right]^{-1} \right) \right). 
\label{eqn:optimal-r}
\eeq

\begin{figure*}
\begin{align}
\label{eqn:modified-dual-problem}
\text{max}_r & \: q = \omega \bigg(\frac{A}{r}\bigg)^{r} \bigg( (\nu - 1)V\bigg)^{\frac{1}{\nu - 1}} \bigg( \frac{R}{1-r}\bigg)^{1-r} \bigg(r + \frac{1}{\nu-1}\bigg)^{r + \frac{1}{\nu-1}} (1-r)^{1-r} \\ 
\text{s.t}. & \quad 0 \leq r \leq 1. 
\label{eqn:bound-on-r}
\end{align}
\hrule
\end{figure*}

Let $q^*$ be the optimal value of the modified dual problem (\ref{eqn:modified-dual-problem})-(\ref{eqn:bound-on-r}). For the optimal primal variables $\lambda^*$ and $S^*$, we have
\begin{align*}
  \omega \lambda^* S^* & = \delta_1^* q^* = q^*, & A (\lambda^*)^{-1} &= \delta_2^* q^* = r^* q^*,   \\
  V (S^{*})^{1-\nu} &= \delta_3^* q^* = \frac{q^{*}}{\nu -1}, & R (\lambda^*)^{-1} &= \delta_4^* q^* = (1-r^*) q^*. 
\end{align*}
By adding the expressions for $A (\lambda^*)^{-1}$ and $R (\lambda^*)^{-1}$, we obtain $A (\lambda^*)^{-1} + R (\lambda^*)^{-1} = q^*$, which we can solve to obtain $\lambda^* = \frac{A+R}{q^*}$. Also, we have $S^* = \left(\frac{V(\nu-1)}{q^*}\right)^{1/(\nu-1)}$. Note that for $S^*$ to be positive, we must have $\nu > 1$. Hence we have the following proposition:

\begin{proposition}
Assuming $A > 0$, $V>0$, and $\nu > 1$, the optimal solution to (P1) is 
\begin{align}
\label{eqn:optlambda_geo}
\lambda^* &= \frac{A+R}{q^*}, \\ 
S^* &= \left(\frac{V(\nu-1)}{q^*}\right)^{1/(\nu-1)},
\label{eqn:optS_geo}
\end{align}
where $q^*$ is the optima of the one dimensional problem (\ref{eqn:modified-dual-problem}) -- (\ref{eqn:bound-on-r}) evaluated at $r^*$ in  (\ref{eqn:optimal-r}).
\label{prop:optSoptlambda_geo}
\end{proposition}


Note that the value of $q$ is indeterminate at $r = 0$ and $r=1$. Since $\lim_{r \rightarrow 1} (1-r)^{1-r} = 1$ and $\lim_{r \rightarrow 1} (\frac{R}{1-r})^{1-r} = 1$, the limiting value of $q^*$ as $r \rightarrow 1$ is
$\lim_{r^* \to 1} q^* = \omega A \left((\nu - 1)V\right)^{1/(\nu - 1)} \left(\frac{\nu}{\nu-1}\right)^{\frac{\nu}{\nu-1}}$. Likewise, since we have the limiting values of $\lim_{r \to  0} (A/r)^r = 1$, the limiting value of $q^*$ as $r \to 0$ is 
$\lim_{r^* \to 0} q^* = \omega R \left((\nu - 1)V\right)^{1/(\nu - 1)} \left(\frac{1}{\nu-1}\right)^{\frac{1}{\nu-1}}$. 


Given the price of infrastructure $\omega$, the optimal strategy of the MNO which is computed by \textbf{Proposition~\ref{prop:optSoptlambda_geo}} will give the minimum amount of cache per unit area, $\lambda^* S^*$, in order to achieve the delay constraint at a typical UE.

\section{Stackelberg Game Model for Pricing: The InP Problem}
\label{subsec:oneInP_oneMNO}

\subsection{Optimization Formulation}

In this section, we consider the InP's strategy when renting out its infrastructure to $K$ MNOs. Due to the infrastructure sharing deployment, both BS and cache storage will be shared by multiple MNOs. Since we consider the MNOs to  request the most popular files, the largest number of files requested to be cached will also accommodate any smaller requests. For example, if MNO-$1$ requests 10 most popular files to be cached and MNO-$2$ requests 15 most popular files, then InP can satisfy both their needs by caching 15 most popular files. This is because the 10 most popular files requested by MNO-$1$ will automatically be included among the 15 most popular files. In general, since $\Scal_k$ is the set of most popular files requested to be cached by the MNO-$k$, we can order the sets $\Scal_k$ as $\mathcal{S}_{\pi(1)} \subseteq \ldots \subseteq \mathcal{S}_{\pi(K)}$, where $\pi$ represents the permutation of set $\Kcal$. Thus, it is sufficient for InP to cache the largest set $\Scal_{\pi(K)}$ to meet every smaller demand. Likewise, using similar reasoning, the InP needs to address only the largest cache intensity, $\max_k \{ \lambda_k S_k \}$, required by the MNOs. Let the maximum of $\lambda_k S_k$ be $\lambda^* S^* = \text{max}_k \{\lambda_k S_k\}$.

We model the pricing problem of the infrastructure as a Stackelberg game, where the InP is assumed to be the leader while the MNOs are assumed to be the followers. However, since the InP only needs to address the largest demand,  the Stackelberg game that the InP needs to solve is in essence a one leader - one follower game. In a Stackelberg game, the leader is assumed to have sufficient foresight to be able to anticipate the followers' strategy.  

The InP aims at maximizing its revenue obtained by leasing the infrastructure to the MNOs, while minimizing the operational expenses in terms of power consumption. Assuming that the InP considers the worst-case scenario where all the MNOs use the same BSs, the transmission power of each BS will be $Kp$. Since the InP is renting out $\lambda^*$ BSs per unit area, the power consumption per unit area (areal power consumption) is then given by $Y(\lambda^*) = \lambda^* (Kp + p_c)$, where $p_c$ denotes a fixed amount of circuit power. 
%
%
%
When the infrastructure is shared by $K$ MNOs, we can formulate the optimization problem of InP as
\begin{align}
\label{eqn:InPleader}
(Q0) \quad \text{max}_{\omega} \quad &\; \omega \lambda^* S^* - \theta Y(\lambda^*),
\end{align}
$\omega$ is the price of cache per unit area, and $\theta$ is the price of areal power consumption, where $\omega, \theta > 0$. Note that, we are not dealing with how other resources, e.g., computing, server, or transmission capacity are shared. We only consider the case where the cache storage in a unit area is shared among the MNOs.

\subsection{Optimal Strategy of the InP}
In order to obtain the solution of the Stackelberg game, we use backward induction method. Accordingly, we first solve the follower subgame problem. This essentially amounts to solving the optimal strategy of the MNO with largest demand, which was covered in the previous section. The follower's solution is then substituted into the leader's subgame problem, after which the leader's problem is solved. The solution to the leader subgame gives the Stackelberg equilibrium.

 For our case, we substitute the $\lambda^*$ and $S^*$ of the follower MNOs from \textbf{Proposition~\ref{prop:optSoptlambda_geo}} to (\ref{eqn:InPleader}). From \textbf{Proposition~\ref{prop:optSoptlambda_geo}}, we can express the optimal values of $S^*$ and $\lambda^*$ in terms of $\omega$ as $\lambda^* = \frac {T}{\omega}$ and $S^* = U\omega^{- \frac{1}{(\nu-1)}}$, where $T = (A+R)/(q^*/\omega)$ and $U = [V(\nu - 1)/(q^*/\omega)]^{1/(\nu-1)}$. Note that in these expressions for $T$ and $U$, from (\ref{eqn:modified-dual-problem}) the term $q^*/\omega$ is independent of $\omega$, making $T$ and $U$ independent of $\omega$ as well. This converts the first term of (\ref{eqn:InPleader}) into
\begin{align*} 
\omega \max_k \{ \lambda_k S_k \}  &= \omega \lambda^* S^* \quad \Longleftrightarrow \quad  \max_k \{ U_k T_k \} \; \omega^{-1/(\nu - 1)} = U T \omega^{-1/(\nu-1)}.
\label{eqn:tot_rent_fromQ0}
\end{align*}
That is, $UT = \max_k\{ U_k T_k \}$. Also, the second term in (\ref{eqn:InPleader}) is converted into
\[ \theta Y(\lambda^*) = \lambda^* \theta (Kp+p_c) = T \bar{p}\omega^{-1}, \]
where $\bar{p} = \theta(Kp+p_c)$.
Therefore, we can rewrite the maximization problem in (\ref{eqn:InPleader}) as an equivalent minimization problem:
\begin{align}
\label{eqn:InPmin}
(Q1)\quad \text{min}_{\omega>0} \quad & T \bar{p} \omega^{-1} - U T \omega^{-1/(\nu-1)}.
\end{align}

The problem (Q1) is a signomial optimization problem over the price variable $\omega$. In general, the problem (Q1) is a non-convex problem however, this problem becomes convex at some values of $\nu$. Nevertheless, we can obtain the solution to (Q1) via successive geometric programming. In order to solve (Q1), let us introduce an auxiliary variable $z\geq0$ such that it upper bounds the objective function (\ref{eqn:InPmin})
\begin{align}
z &\geq T \bar{p} \omega^{-1} - U T \omega^{-1/(\nu-1)}.
\label{eqn:upper-bound-obj-fun-Q1}
\end{align}

Since minimizing the upper bound $z$ minimizes the objective function (\ref{eqn:InPmin}) as well, the problem (Q1) can be equivalently re-written in terms of this auxiliary variable as
\begin{align}
\label{eqn:InPmin_transform}
(Q2)\quad \chi &= \text{min}_{\omega>0} \: z \\
\text{s.t.} \quad & \frac{T \bar{p}\omega^{-1}}{z + U T \omega^{-1/(\nu-1)}} \leq 1. \label{eqn:transform_constraint} 
\end{align}
Here, the constraint (\ref{eqn:transform_constraint}) is obtained after some algebraic manipulation of the bound in (\ref{eqn:upper-bound-obj-fun-Q1}). 

Since the constraint (\ref{eqn:transform_constraint}) is a ratio of two posynomials, the problem (Q2) is also referred to as a complementary geometric programming problem. Following the approach outlined in~\cite{Avriel1971,Morris1972}, we can approximate (Q2) by a suitable geometric program by substituting the posynomial in the denominator of (\ref{eqn:transform_constraint}) by a monomial. We can transform a posynomial into a monomial using the arithmetic-geometric mean inequality $\sum_i x_i \geq \prod_i \left( \frac{x_i}{w_i} \right)^{w_i}$ for non-negative numbers $x_i \geq 0$, 
where $\sum_i w_i = 1$ and equality if and only if all $x_i /w_i$ are the same.
To see this, let us denote the posynomial in the denominator of (\ref{eqn:transform_constraint}) as $Q(z,\omega) = z + U T \omega^{-1/(\nu-1)},$ and its evaluation at point $(\bar{z},\bar{\omega})$ as 
$\bar{Q} = Q(\bar{z},\bar{\omega}) = \bar{z} + U T \bar{\omega}^{-\frac{1}{\nu -1}}.$
Substituting each terms of $Q(z,\omega)$ into $x_i$ and since $(\bar{z} + U T \bar{\omega}^{-\frac{1}{\nu -1}})/\bar{Q} = 1$, we have the lower bound from the arithmetic-geometric mean inequality as $Q(z,\omega) \geq Q(z, \omega, \bar{z}, \bar{\omega})$, where $Q(z, \omega, \bar{z}, \bar{\omega})$ is a monomial given by
\begin{align*} 
Q(z, \omega, \bar{z}, \bar{\omega}) &= \left( \frac{ z \bar{Q}} {\bar{z}} \right)^{\frac{\bar{z}}{\bar{Q}}} \left( \frac{ \omega^\frac{-1}{\nu-1} \bar{Q}}{ \bar{\omega}^{\frac{-1}{\nu-1}}} \right)^{\frac{U T\bar{\omega}^{\frac{-1}{\nu-1}}}{\bar{Q}}} = E z^{\bar{\alpha}} \omega^{\bar{\beta}}. 
\end{align*}
The parameters are given by,
\begin{align}
\bar{\alpha} = \frac{\bar{z}}{\bar{Q}}, \qquad  \bar{\beta} = \frac{- U T \bar{\omega}^{\frac{-1}{\nu-1}}}{(\nu-1) \bar{Q}}, \qquad E = \frac{\bar{Q}}{\bar{z}^{\bar{\alpha}} \bar{\omega}^{\bar{\beta}}}.
\label{eqn:parameters-SGA}
\end{align}
Therefore, the problem (Q2) can be transformed into an approximate geometric program by approximating the denominator of (\ref{eqn:transform_constraint}) by the lower bound $Q(z, \omega, \bar{z}, \bar{\omega})$ as follows:
\begin{align}
\label{eqn:InP_geo}
(Q3) \quad \chi &= \text{min}_{\omega} \: z \\
\text{s.t.} \quad & \frac{T\bar{p}\omega^{-1}}{E z^{\bar{\alpha}} \omega^{\bar{\beta}} } \leq 1. 
\end{align}
The problem (Q3) is now a geometric program which can be solved analytically. Note that the degree of difficulty of this problem is zero. The dual maximization problem of (Q3) is given by 
\begin{align} 
\text{max}_{\boldsymbol{\bar{\delta}}} \: \chi_d =& \left(\frac{1}{\bar{\delta}_1} \right)^{\bar{\delta}_1}\left( \frac{T \bar{p}}{\bar{\delta}_2 E} \right)^{\bar{\delta}_2} \bar{\delta}_2^{\bar{\delta}_2}, \\
\text{s.t.} \quad & \bar{\delta}_1 = 1, \label{eqn:normality_InP}\\
 & \begin{pmatrix}
         0 & -1-\bar{\beta} \\
         1 & -\bar{\alpha} 
        \end{pmatrix}   
        \begin{pmatrix}
        \bar{\delta}_1 \\ \bar{\delta}_2 
         \end{pmatrix}       
      = 0.
\label{eqn:geo_orthogonality_InP}
\end{align} 
We can solve for the optimal values of dual variables $\boldsymbol{\bar{\delta}}$ directly from (\ref{eqn:normality_InP}) and (\ref{eqn:geo_orthogonality_InP}). The matrix multiplication from (\ref{eqn:geo_orthogonality_InP}) gives $(-1- \bar{\beta})\bar{\delta}_2 = 0$ and $\bar{\delta}_1 - \bar{\alpha} \bar{\delta}_2 = 0$. Summing these two equations give $(-1- \bar{\beta})\bar{\delta}_2 + \bar{\delta}_1 - \bar{\alpha} \bar{\delta}_2 = 0$. Since $\bar{\delta}_1 = 1$, we obtain $\bar{\delta}_2 = 1/(1+\bar{\alpha}+\bar{\beta})$.
By substituting the values of $\bar{\delta}_1$ and $\bar{\delta}_2$ in the dual problem $\chi_d$, the optimal dual function $\chi_d^*$ yields
\begin{align}
\chi_d^* &= \left( \frac{T \bar{p}}{E}\right)^{\frac{1}{1+\bar{\alpha}+\bar{\beta}}}.
\label{eqn:dual-objective-SGA}
\end{align}
In order to find the primal variables $z^*$ and $\omega^*$ from $\chi_d^*$, we have
\[z^* = \bar{\delta}_1^* \chi_d^* = \left( \frac{T \bar{p}}{E}\right)^{\frac{1}{1+\bar{\alpha}+\bar{\beta}}},\] 
and $\left( \frac{T \bar{p} (\omega^*)^{-1}}{E (z^*)^{\bar{\alpha}} (\omega^*)^{\bar{\beta}}} \right) = \delta_2^* \chi_d^* = \left( \frac{1}{1 + \bar{\alpha} + \bar{\beta}} \right) \left( \frac{T \bar{p}}{E} \right)^{\frac{1}{1+\bar{\alpha} + \bar{\beta}}}.$
As such, we can compute 
\[\omega^* = \left( \frac{T \bar{p} (1 + \bar{\alpha} + \bar{\beta})}{E (z^*)^{\bar{\alpha}}} \left( \frac{E}{T \bar{p}}\right)^{1/(1 + \bar{\alpha} + \bar{\beta})} \right)^{1/(1+\bar{\beta})}.\] 
Since this is the first approximate values of $z^*$ and $\omega^*$, we can substitute these values back into $Q(z,\omega,z^*,\omega^*)$ and repeat the process, leading to new values of to $z^*$ and $\omega^*$. Thus, we have a successive geometric programming algorithm as given in \tbf{Algorithm~\ref{alg:SGA_Algorithm}}. \tbf{Algorithm~\ref{alg:SGA_Algorithm}} converges to a local optima of (Q2)~\cite{Avriel1970}. This leads to the following proposition:

\begin{proposition}
\textbf{Algorithm~\ref{alg:SGA_Algorithm}} converges to a locally optimal solution of the primal problem (Q0) for the InP. This gives the Stackelberg equilibrium of the one leader and one follower game. 
\label{prop:Stacklerberg_Equi}
\end{proposition}

\begin{algorithm}[t]
\caption{Successive Geometric Programming}
\label{alg:SGA_Algorithm}
 \begin{algorithmic}[1]
 \STATE Set $z(0)$ and $\omega(0)$ to any arbitrary feasible values 
 \REPEAT
 \STATE Calculate $\bar{\alpha}(t), \bar{\beta}(t), E(t)$ using (\ref{eqn:parameters-SGA}) 
 \STATE Calculate $\chi_d(t)$ using (\ref{eqn:dual-objective-SGA})
 \STATE $\omega(t+1) \leftarrow\left( \frac{T \bar{p} (1 + \bar{\alpha} + \bar{\beta})}{E (z^*)^{\bar{\alpha}}} \left( \frac{E}{T \bar{p}}\right)^{1/(1 + \bar{\alpha} + \bar{\beta})} \right)^{1/(1+\bar{\beta})}$
\STATE $z(t+1) \leftarrow \chi_d(t)$ 
\UNTIL convergence
 \end{algorithmic}
 \end{algorithm}

\section{Cost Sharing Among MNOs}
\label{subsec:coalition_shapleyvalue}
After an InP computes the rent of the infrastructure so as to maximize its revenue from the largest demand of an MNO, the InP will declare the total rent to all MNOs. The MNOs will cooperate with each other and find a way to fairly divide the infrastructure rental among themselves. Therefore, we model this situation by a cooperative game for rent sharing using Shapley value. The coalition form of $K$-person game is given by $(\Kcal, v)$, where $\Kcal$, the set of $K$ MNOs, is the set of players. The characteristic function of the game is denoted by $v$, where $v : 2^K \rightarrow \mathbb{R}$. The characteristic function maps every $2^K$ possible coalitions to a real number, referred to as the value of the coalition. When all $K$ MNOs cooperate among themselves and form a single group, it is called the ``grand coalition''. The value of the grand coalition is $v(\Kcal)$. For the case when a subset of MNOs cooperate with each other and form a coalition $\mathcal{C} \subseteq \Kcal$, we define the characteristic function of $\mathcal{C}$ as
\beq 
v(\emptyset) = 0 \quad \mathrm{and} \quad v(\mathcal{C}) = \text{max}_{k \in \mathcal{C}} \; \omega^* \lambda_k^* S_k^*,
\label{eqn:characteristic-fun}
\eeq
where $\omega^*$ is the optimal price set by the InP following \textbf{Proposition~\ref{prop:Stacklerberg_Equi}}. 
\begin{proposition}
The characteristic function $v$ given in (\ref{eqn:characteristic-fun}) satisfies sub-additivity property, $v(\mathcal{C}_1 \cup \mathcal{C}_2) \leq v(\mathcal{C}_1) + v(\mathcal{C}_2)$, for any two coalitions $\Ccal_1$ and $\Ccal_2$, where $\mathcal{C}_1 \cap \mathcal{C}_2 = \emptyset$. 
\end{proposition}
\begin{IEEEproof}
Let $k^* = \argmax_{k \in \Ccal_1 \cup \Ccal_2} \omega^*\lambda_k^* S_k^*$. Since $\Ccal_1$ and $\Ccal_2$ are disjoint, $k^*$ must belong to either $\Ccal_1$ or $\Ccal_2$, but not both. If $k^* \in \Ccal_1$, we have  $v(\Ccal_1 \cup \Ccal_2) = v(\Ccal_1)$. Therefore, $v(\Ccal_1 \cup \Ccal_2) \leq  v(\Ccal_1) + v(\Ccal_2)$. Likewise, if $k^* \in \Ccal_2$, then $v(\Ccal_1 \cup \Ccal_2) = v(\Ccal_2)$. Again, $v(\Ccal_1 \cup \Ccal_2) \leq  v(\Ccal_1) + v(\Ccal_2)$. Thus, $v$ given in (\ref{eqn:characteristic-fun}) is sub-additive.
\end{IEEEproof}

Note that the sub-additivity implies that the MNOs forming a bigger coalition will have smaller cost. Thus, the MNOs have an incentive to form larger groups. Let the cost allocated to MNO-$k$ be $\psi_k$. Then, the value of the coalition $\mathcal{C}$ should be divided among each MNO comprising the coalition $\Ccal$ such that $v(\mathcal{C}) = \sum_{k \in \mathcal{C}} \psi_k$.

We select the Shapley value as the cost allocation strategy, since it does not rely on the super-additivity or sub-additivity of the characteristic function~\cite{Myerson1991}. A desirable property of the Shapley value is that it can be used as a solution of a smaller game with only a subset of players taken from the grand coalition~\cite{Aumann1974}. 
The Shapley value function, $\psi$, is a function that assigns to each possible characteristic function of a $K$ MNO game, $v$ with a $K$-tuple, $\psi(v) = (\psi_1(v), \psi_2(v), \ldots, \psi_K(v))$. The $\psi_k(v)$ represents the worth or value of MNO-$k$ in the game with characteristic function $v$ and is defined by the following axioms of fairness:
\begin{enumerate}
\item Efficiency: $\sum_{k \in \Kcal} \psi_k(v) = v(\Kcal)$.
\item Symmetry: If $k$ and $l$ are such that $v(\mathcal{C} \cup \{k\}) = v(\mathcal{C} \cup \{l\})$ for every coalition $\mathcal{C}$ not containing $k$ and $l$, then $\psi_k(v) = \psi_l(v)$. 
\item Dummy: If $i$ is such that $v(\mathcal{C}) = v(\mathcal{C} \cup \{i\})$ for every coalition $\Ccal$ not containing $i$, then $\psi_i(v) = 0$.
\item Additivity: If $u$ and $v$ are characteristic functions, then $\psi(u + v) = \psi(u) + \psi(v)$. 
\end{enumerate}
There exists a unique function that satisfies all these fairness axioms which is given by:
\beq
\psi_k(v) = \sum\limits_{\substack{\mathcal{C} \subseteq \Kcal \\ k \in \mathcal{C}}} \frac{ (|\mathcal{C}| -1)!(n - |\mathcal{C}|)!}{k!} [v(\mathcal{C}) - v(\mathcal{C} - \{k\})].
\label{eqn:def-shapely-value}
\eeq
This gives the average marginal contribution made by $k$-th MNO when it joins a random coalition $\mathcal{C}$. We take this value as the fair payoff allocation among the MNOs inside the coalition. 

The direct computation of the Shapely value using the analytical formula given in (\ref{eqn:def-shapely-value}) quickly becomes computationally infeasible as the number of MNOs increases. However, due to the peculiar characteristic function for our rent sharing problem, as given in (\ref{eqn:characteristic-fun}), we can find a simple algorithm to allocate the cost among the MNOs by recognizing that the our problem is equivalent to the airport runway cost sharing problem studied by Littlechild and Owens~\cite{Littlechild1973}. For our case, the total rent, $\omega^* \lambda^* S^*$, as obtained by \textbf{Algorithm~\ref{alg:SGA_Algorithm}}, is set by the InP; and the MNOs try to divide the rent among each other according to their required cache intensities $\lambda_k^* S_k^*$ as obtained from \textbf{Proposition~\ref{prop:optSoptlambda_geo}}. 

\textbf{Algorithm~\ref{alg:Airport_Runway_CostSharing}} gives the rent sharing algorithm among MNOs. Depending on the requirement of the MNOs, the MNOs are first arranged in an ascending order of their cache intensities $\lambda_{\pi(1)}^* S_{\pi(1)}^* \leq \ldots \leq \lambda_{\pi(K)}^* S_{\pi(K)}^*$, where $\pi$ denotes the permutation of set $\Kcal$. The cost of catering to the smallest cache intensity, $\omega^* \lambda^*_{\pi(1)} S^*_{\pi(1)}$, is divided equally among all $K$ MNOs. Then, the incremental cost $\Delta = \omega^* \lambda^*_{\pi(2)} S^*_{\pi(2)} - \omega^* \lambda^*_{\pi(1)} S^*_{\pi(1)}$ of catering to the second smallest demand is shared equally among all the MNOs, except the MNO with smallest demand. The process is continued until the incremental cost $\Delta = \omega^* \lambda^*_{\pi(K)} S^*_{\pi(K)} - \omega^* \lambda^*_{\pi(K-1)} S^*_{\pi(K-1)}$ is allocated only to the MNO with largest demand. The cost allocated among the MNOs by the \textbf{Algorithm~\ref{alg:Airport_Runway_CostSharing}} is equivalent to the Shapley value of the coalition game $(\Kcal,v)$~\cite{Littlechild1973}. Note that \textbf{Algorithm~\ref{alg:Airport_Runway_CostSharing}} has the worst-case computational complexity of $O(K^2)$. 

\begin{algorithm}[t]
\caption{Cost sharing among MNOs}
\label{alg:Airport_Runway_CostSharing}
 \begin{algorithmic}[1]
 \STATE Initialize $\lambda_0 S_0 = 0$ and $\psi_k = 0$ for all $k\in\Kcal$
 \STATE Arrange the MNOs in an ascending order of their cache intensity $\lambda_{\pi(1)}^* S_{\pi(1)}^* \leq \ldots \leq \lambda_{\pi(K)}^* S_{\pi(K)}^*$, where $\pi$ is the permutation of set $\Kcal$
 \FOR{$k = 1$ \TO $K$} 
 \STATE $\Delta = \omega^* \lambda^*_{\pi(k)} S^*_{\pi(k)} - \omega^* \lambda^*_{\pi(k-1)} S^*_{\pi(k-1)}$.
 \FOR{$i = k$ \TO $K$}
 \STATE $\psi_{\pi(i)} \leftarrow \psi_{\pi(i)} + \Delta/(K-k+1)$ 
\ENDFOR
\ENDFOR
 \end{algorithmic}
 \end{algorithm}

\section{Numerical Results}
\label{subsec:results}
The baseline setting of simulation environment is as follows : the transmit power of BS is $p = 1$ Watt, noise power is $\sigma^2 = -150$ dBm, user intensity is $\xi = 60/(\pi \times 500^2)$, number of video files in the cloud is $F = 10^5$; size of the file requested by each UE is $x_f = 10^9$ bits, path-loss exponent is $\alpha = 5$, i.e., suburban area without line of sight \cite{pathloss}. The SINR threshold is $\bar{T} = 10$ dB, the probability that UE requests service from BS is $\eta = 0.014$~\cite{Anastasios2013}. Each MNO is assumed to have the same number of subband as $L =6$. Based on $5$G requirements in \cite{5GPPP}, the delay of data transmission should be less than $10^{-3}$ secs. Therefore, we try to constrain the total delay such that $\text{Pr}(D \geq 10^{-3}) \leq 0.1$, where $D_{\text{th}} = 10^{-3}$ sec, and $\gamma = 0.1$ in (\ref{eqn:geometric_prog}). We assume that there is a single server in the cloud, where the mean arrival rate of file requests is $\phi = 0.8$, mean service time $\tau = 5 \times 10^{-3}$, the coefficient of variation of inter-arrival time and service time are taken to be $c_a = 2$ and $c_s = 1$ as such, $\Ebb[D_{\text{bh}}] = 0.0051$ sec.

\subsection{Optimal Strategy of an MNO from \textbf{Proposition~\ref{prop:optSoptlambda_geo}}}
\label{subsec:opt-strategy-MNO}
\begin{figure}[h]
\begin{minipage}{0.48\textwidth}
\begin{center}
	\includegraphics[width=\textwidth]{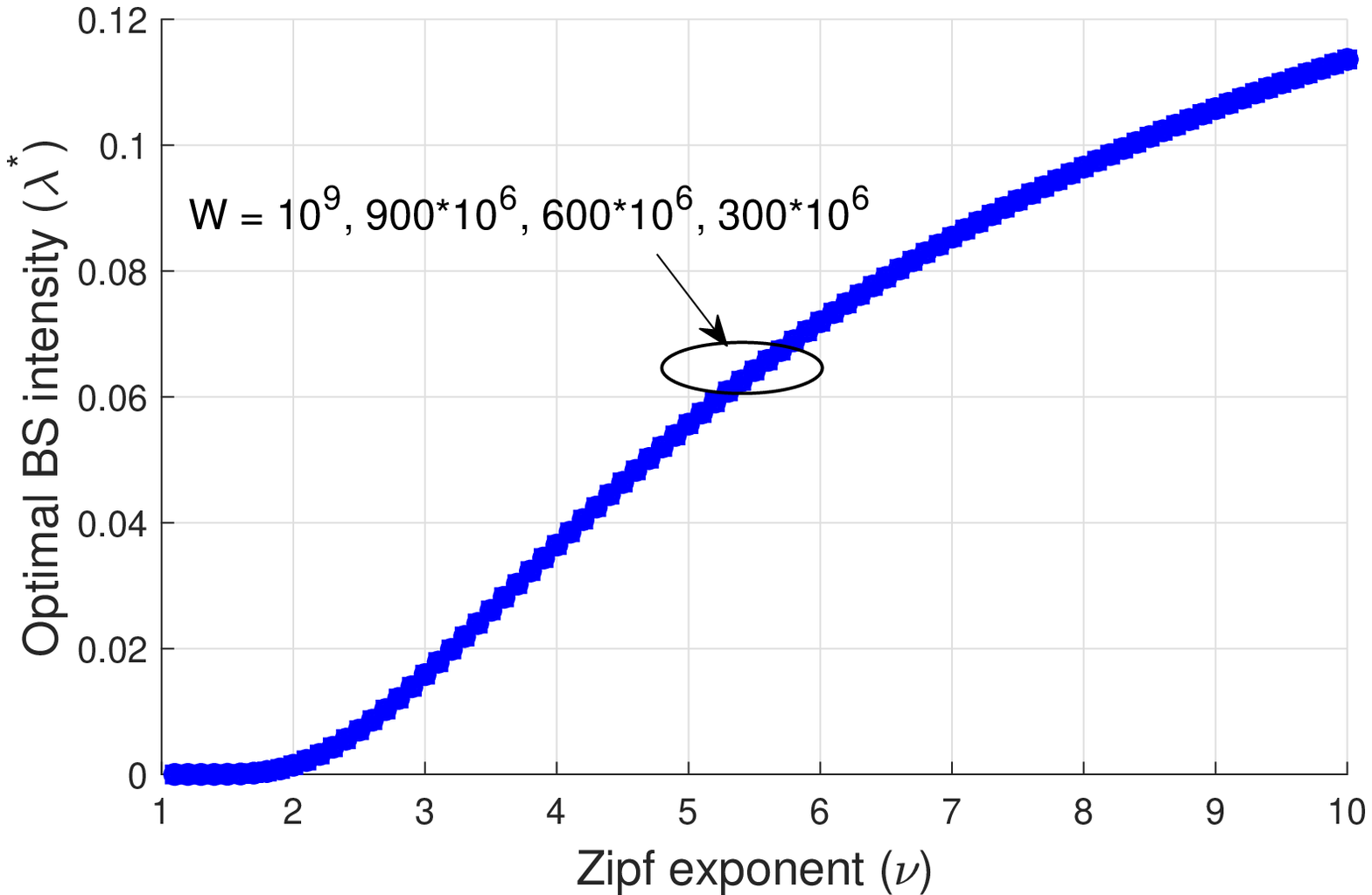}
	\caption{Optimal base station intensity ($\lambda^*$) versus Zipf exponent ($\nu$).}
         \label{fig:Optlambda_nu}
 \end{center}
 \end{minipage}
\hfill
\begin{minipage}{0.48\textwidth}
\begin{center}
	\includegraphics[width=\textwidth]{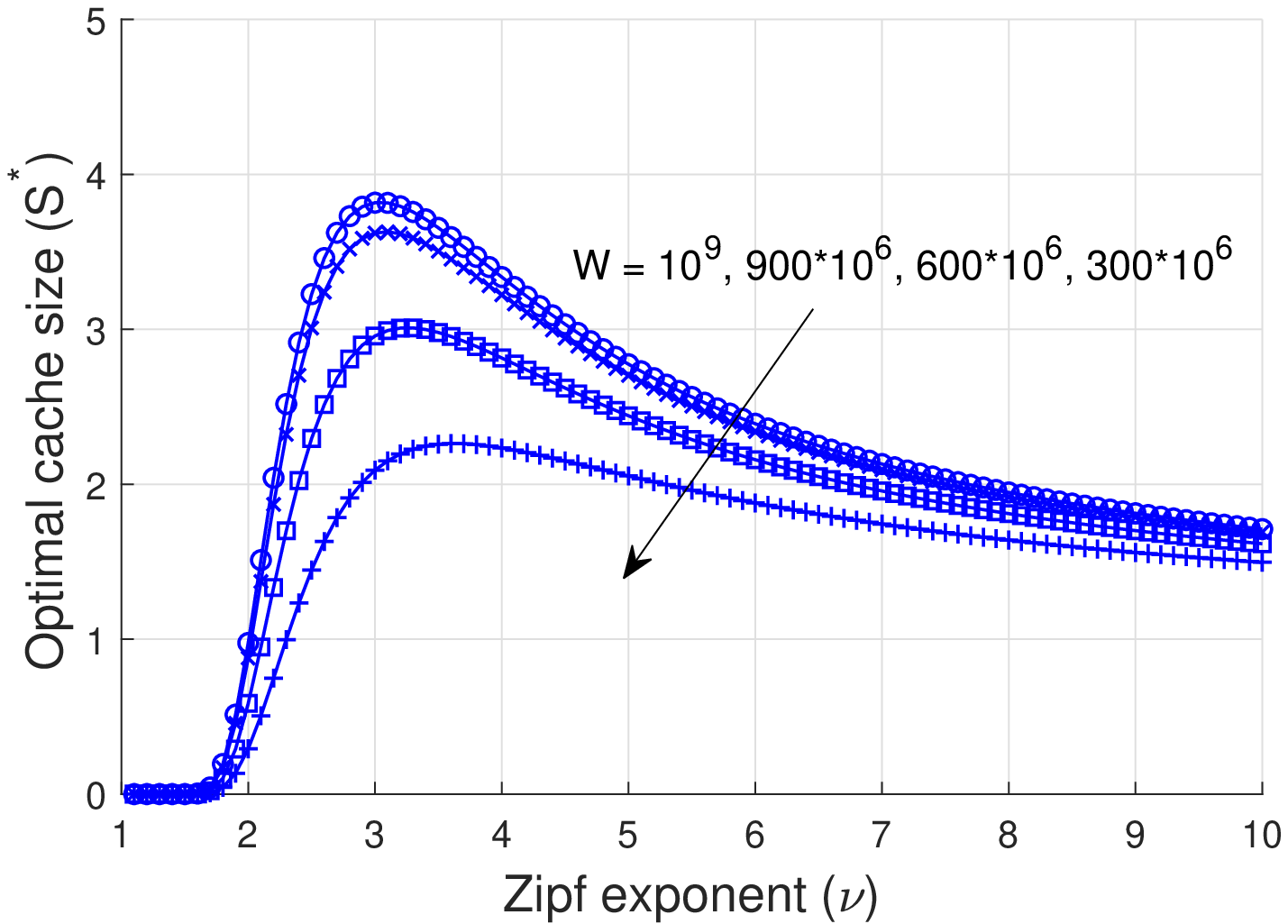}
	\caption{Optimal base station intensity ($\lambda^*$) versus Zipf exponent ($\nu$).}
        \label{fig:OptS_nu_BW}
\end{center}
\end{minipage}
\end{figure}

\begin{figure}[h]
\centering
\includegraphics[height=3 in, width=3 in, keepaspectratio = true]{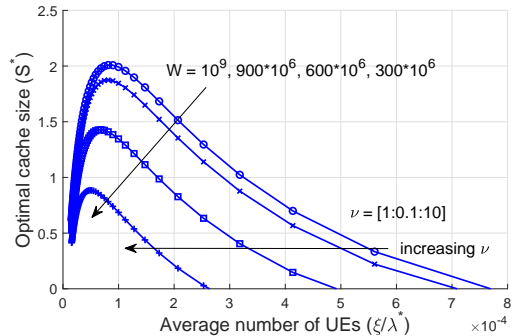}
\caption{Optimal cache size versus average number of UEs per BS ($\xi/\lambda^*$) with varying channel bandwidth ($W$).}
\label{fig:OptCache_AvgUE_BW}
\end{figure}

In Fig.~\ref{fig:Optlambda_nu} and Fig.~\ref{fig:OptS_nu_BW}, we plot the optimal strategy of a single MNO from \textbf{Proposition~\ref{prop:optSoptlambda_geo}}. The optimal BS intensity, $\lambda^*$, and the optimal cache size, $S^*$, versus Zipf exponent, $\nu$, are illustrated in Fig.~\ref{fig:Optlambda_nu} and Fig.~\ref{fig:OptS_nu_BW}, respectively. Also, the optimal cache size, $S^*$, versus average number of UEs per BS, $\xi/\lambda^*$, is shown in Fig.~\ref{fig:OptCache_AvgUE_BW}. The bandwidth ($W$) is assumed to be $W = 10^9, 900\times10^6, 600\times10^6,$ and $300\times10^6$ Hz while the price of cache intensity is $\omega = 10$.

%
In Fig.~\ref{fig:Optlambda_nu}, we see that $\lambda^*$ increases when $\nu$ is increased; however, $\lambda^*$ does not change with $W$ for a given $\nu$. On the other hand, in Fig.~\ref{fig:OptS_nu_BW}, the optimal cache size $S^*$ decreases when $W$ is decreased for a given $\nu$. This is because, for these parameters, we have $r^* = 1$. Also, since a higher bandwidth yields a greater throughput, from $G \propto W$ in (\ref{eqn:goodput}), and since the constant $A \propto C_2 \propto 1/G$, we have $q^* \propto A \propto 1/G$ for the case when $r^*=1$. Thus, from (\ref{eqn:optS_geo}), we have $S^* \propto (1/q^*)^{1/(\nu-1)} \propto G^{1/(\nu-1)} \propto W^{1/(\nu-1)}$. Hence, for $\nu>1$, increasing $W$ leads to a decrease in the cost $q^*$, which in turn leads to an increased cache size $S^*$. On the other hand, since $R \propto 1/G$, from (\ref{eqn:optlambda_geo}) the $G$ term cancels out in the expression for $\lambda^*$, making $\lambda^*$ independent of $W$. For increasing $\nu$, the number of files to be stored in the cache becomes smaller since a greater $\nu$ means more skewness of file popularity. Therefore, while the optimal cache size $S^*$ decreases with an increasing of $\nu$, the BS intensity $\lambda^*$ increases. Subsequently, Fig.~\ref{fig:Optlambda_nu} and Fig.~\ref{fig:OptS_nu_BW} show the tradeoff between $\lambda^*$ and $S^*$ for a given $\nu$.

We vary the values of $\nu$ to be from $1$ to $10$ in each curve in Fig.~\ref{fig:OptCache_AvgUE_BW}. By increasing $\nu$, $\lambda^*$ tends to increase,  hence the average number of UEs $\xi$ will decrease. This results in a  decrease in the average number of UEs per BS, $\xi/\lambda^*$. Also, as $\nu$ increases, the cache amount will first increase and then decrease.

\subsection{Optimal Price ($\omega^*$) and Maximum Profit ($z^*$) of the InP at Stackelberg Equilibrium}
\begin{figure}[h]
\begin{minipage}{0.48\textwidth}
\begin{center}
	\includegraphics[width=\textwidth]{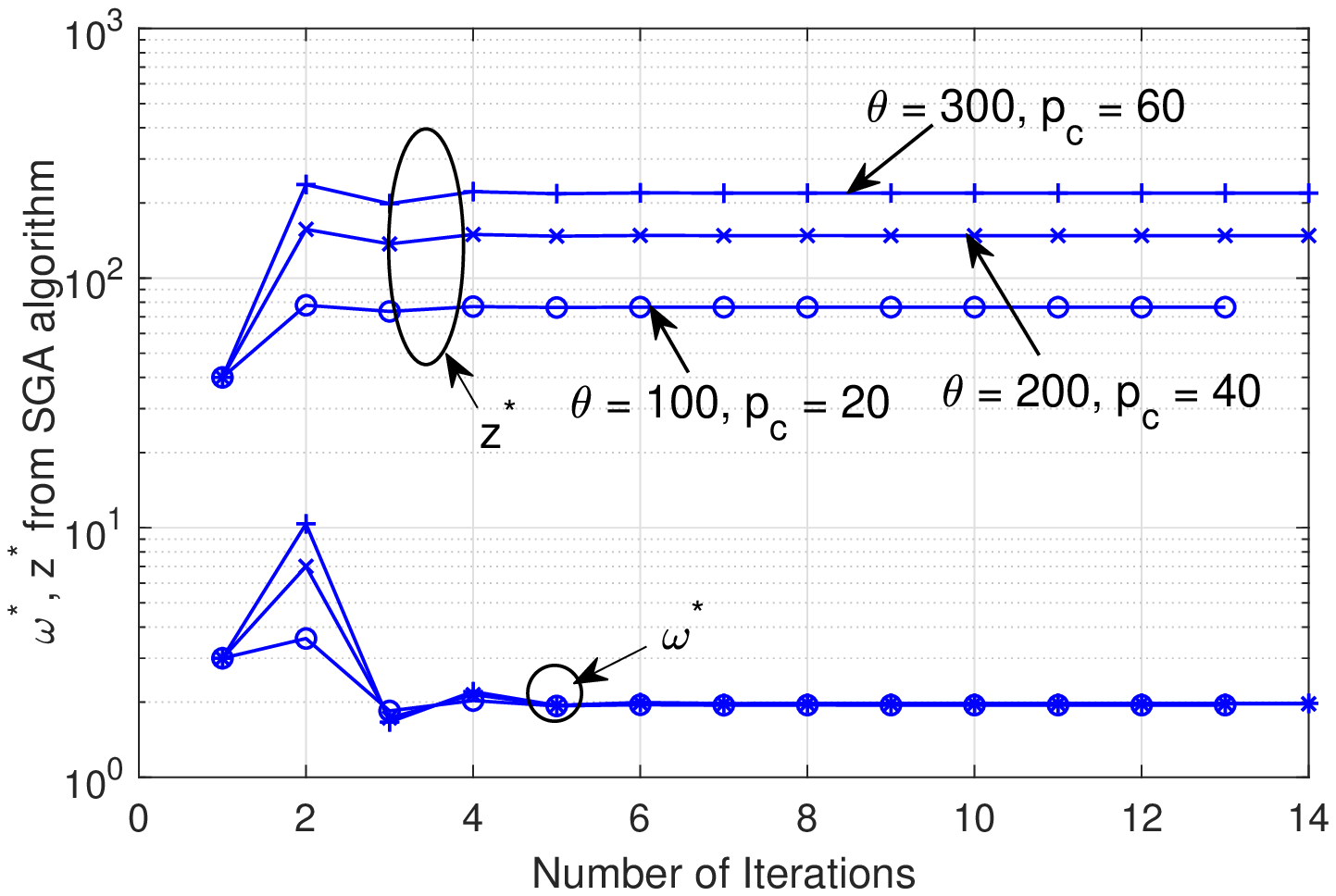}
\caption{Convergence of the optimal price of infrastructure ($\omega^*$) and maximum profit ($z^*$) from \textbf{Algorithm~\ref{alg:SGA_Algorithm}}.}
\label{fig:Converge_Optomega_optz_SGA}
 \end{center}
 \end{minipage}
\hfill
\begin{minipage}{0.48\textwidth}
\begin{center}
	\includegraphics[width=\textwidth]{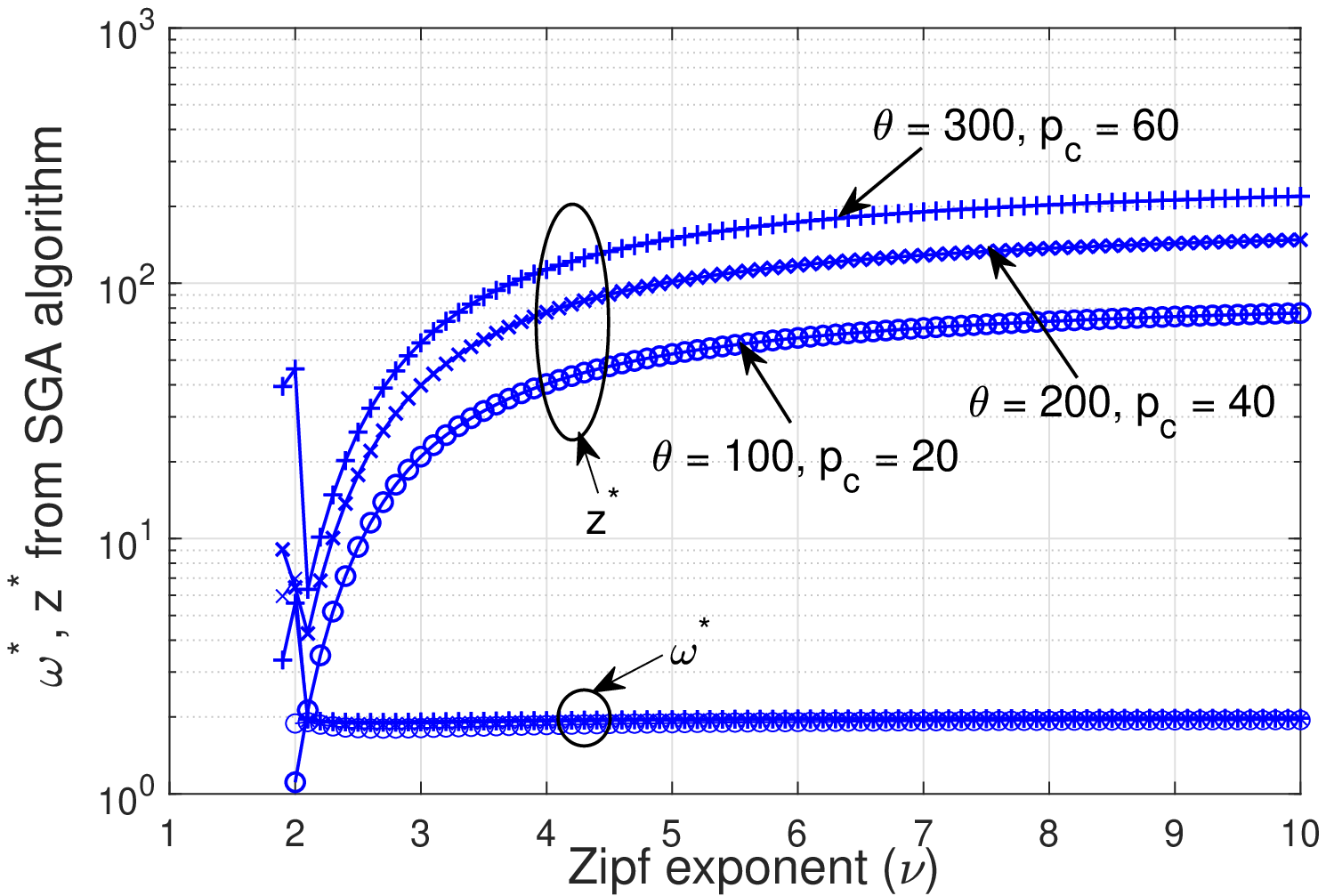}
\caption{Optimal price of infrastructure ($\omega^*$) and maximum profit ($z^*$) from \textbf{Algorithm~\ref{alg:SGA_Algorithm}} versus Zipf exponent ($\nu$).}
\label{fig:Optomega_optz_versus_nu_SGA}
\end{center}
\end{minipage}
\end{figure}

\begin{figure}[h]
\begin{minipage}{0.48\textwidth}
\begin{center}
	\includegraphics[width=\textwidth]{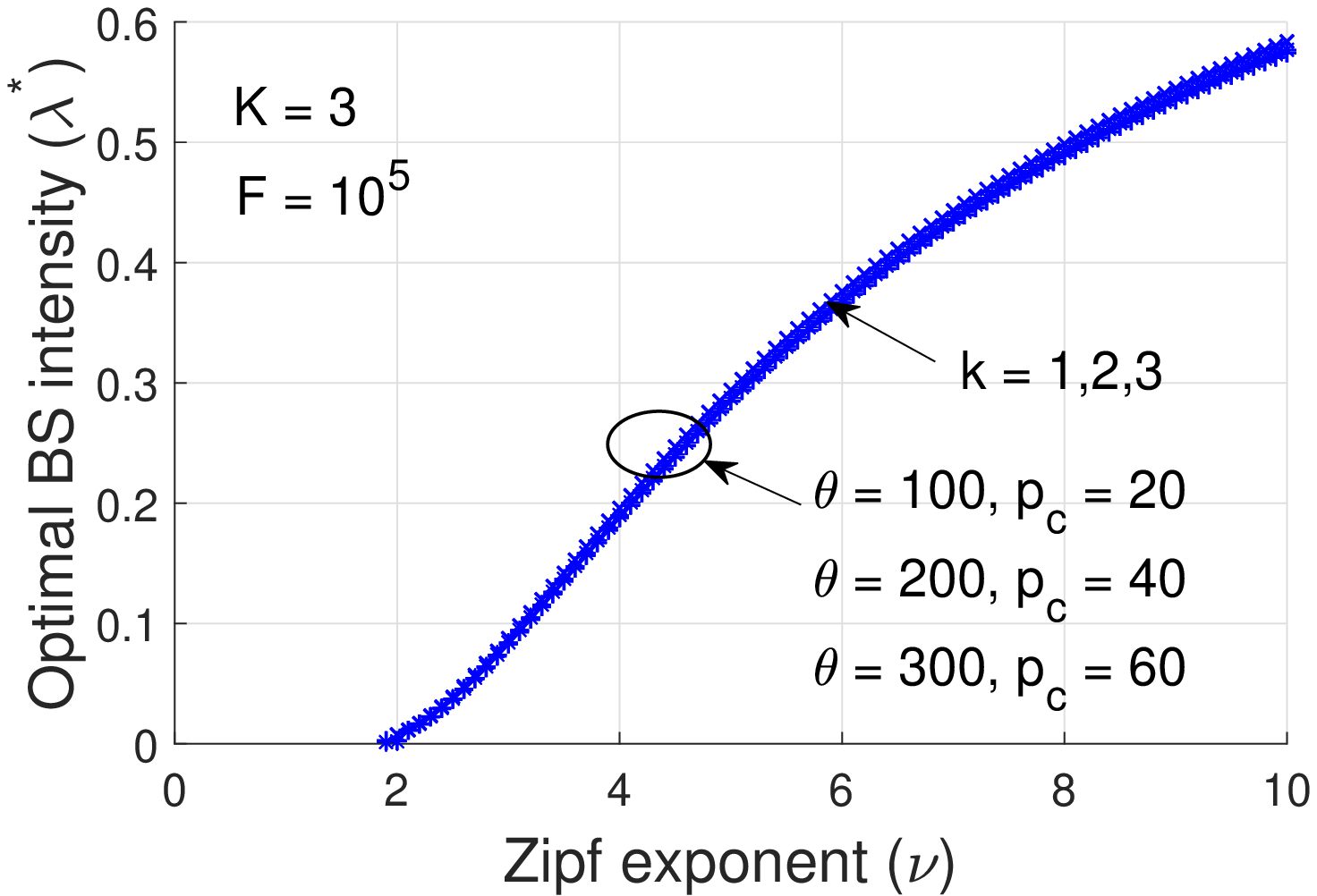}
        \caption{Optimal base station intensity ($\lambda^*$) versus Zipf exponent ($\nu$) at Stackelberg equilibrium.}
      \label{fig:Optlambda_versus_nu_Stackleberg}
     \end{center}
\end{minipage}
\hfill
\begin{minipage}{0.48\textwidth}
\begin{center}
	\includegraphics[width=\textwidth]{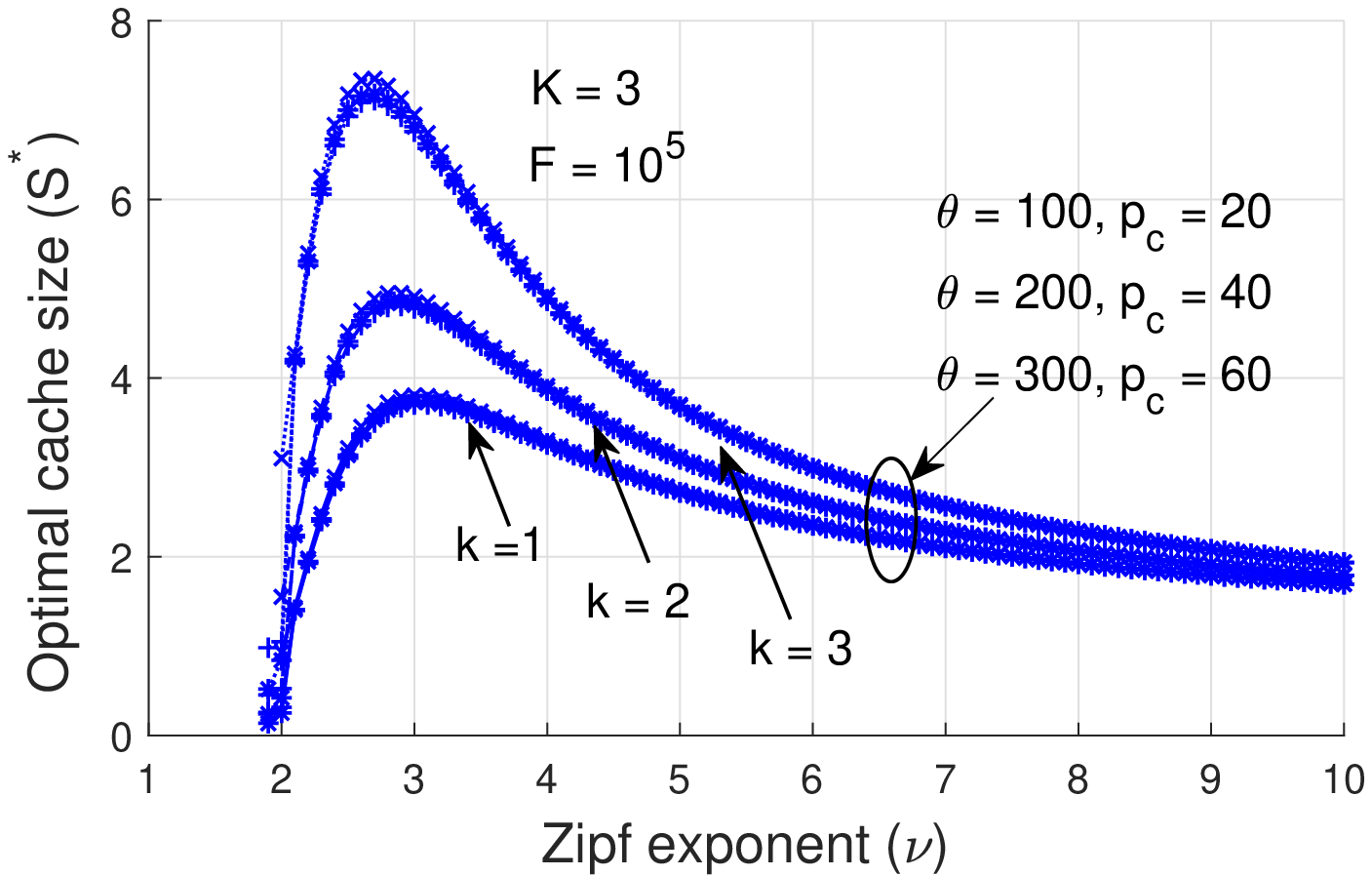}
        \caption{Optimal cache size ($S^*$) Zipf exponent ($\nu$) with ($\omega^*$) from \textbf{Algorithm~\ref{alg:SGA_Algorithm}}. }
     \label{fig:OptS_versus_nu_Stackleberg}\end{center}
\end{minipage}
\end{figure}

After computing the optimal strategy from each follower MNO from \textbf{Proposition~\ref{prop:optSoptlambda_geo}}, due to infrastructure sharing deployment, the InP chooses the largest demand required by the MNOs. The InP will then compute the optimal price of an infrastructure so as to maximize it's revenue using (Q0). In  Figs.~\ref{fig:Converge_Optomega_optz_SGA}-Fig.~\ref{fig:OptS_versus_nu_Stackleberg}, we demonstrate the optimal strategy of the leader InP and the optimal strategy of multiple follower MNOs at Stackelberg equilibrium. We assume that there are three follower MNOs. The optimal price $\omega^*$ and the maximum profit $z^*$ at the equilibrium are given by the successive geometric programming  in \textbf{Algorithm~\ref{alg:SGA_Algorithm}}. The optimal strategies of the MNOs at the Stackelberg equilibrium are shown in Fig.~\ref{fig:Optlambda_versus_nu_Stackleberg} and Fig.~\ref{fig:OptS_versus_nu_Stackleberg}. We illustrate the results assuming the bandwidths $W_k$ of the MNOs to be $[W_1, W_2, W_3]= [300\times10^6, 500\times10^6,1\times10^9]$ Hz.

In Fig.~\ref{fig:Converge_Optomega_optz_SGA}, we show the convergence of the optimal price of infrastructure, $\omega^*$, and the maximum profit, $z^*$. When the price of areal power consumption, $\theta$, and the circuit power, $p_c$, increase, $z^*$ also increases, while both $\theta$ and $p_c$ have very small effect on $\omega^*$. The SGA algorithm is an effective algorithm since both $\omega^*$ and $z^*$ converge within a few number of iterations. In  Fig.~\ref{fig:Optomega_optz_versus_nu_SGA}, $\omega^*$ and $z^*$ are plotted against $\nu$. We can see that when $\nu$ increases, $z^*$ also increases. However, with increasing $\nu$, the $\omega^*$ remains constant. The optimal $\lambda^*$ and the optimal $S^*$ of each MNO at the equilibrium are shown in Fig.~\ref{fig:Optlambda_versus_nu_Stackleberg} and Fig.~\ref{fig:OptS_versus_nu_Stackleberg}, respectively.

In Fig.~\ref{fig:Optlambda_versus_nu_Stackleberg}, the curves for  $\lambda^*$ of each MNO is the same as each other. The trends in Fig.~\ref{fig:Optlambda_versus_nu_Stackleberg} are also very similar to those in Fig.~\ref{fig:Optlambda_nu}. Also, varying of price of infrastructure $\theta$ and circuit power $p_c$ does not have any impact on $\lambda^*$. In Fig.~\ref{fig:OptS_versus_nu_Stackleberg}, we plot $S^*$ of different MNOs versus $\nu$. When the value of $\nu$ changes, the optimal cache size $S^*$ of MNO-$1$ is the smallest while that of  MNO-$3$ is the highest. It is due to the fact that MNO-$3$ has the highest bandwidth, while MNO-$1$ has the lowest bandwidth. Varying the price of areal power consumption $\theta$ and circuit power $p_c$ does not change $S^*$ as well.

\subsection{Maximum Profit of the InP at Stackelberg Equilibrium and Shapley Value}

\begin{figure}[h]
\begin{minipage}{0.48\textwidth}
\begin{center}
	\includegraphics[width=\textwidth]{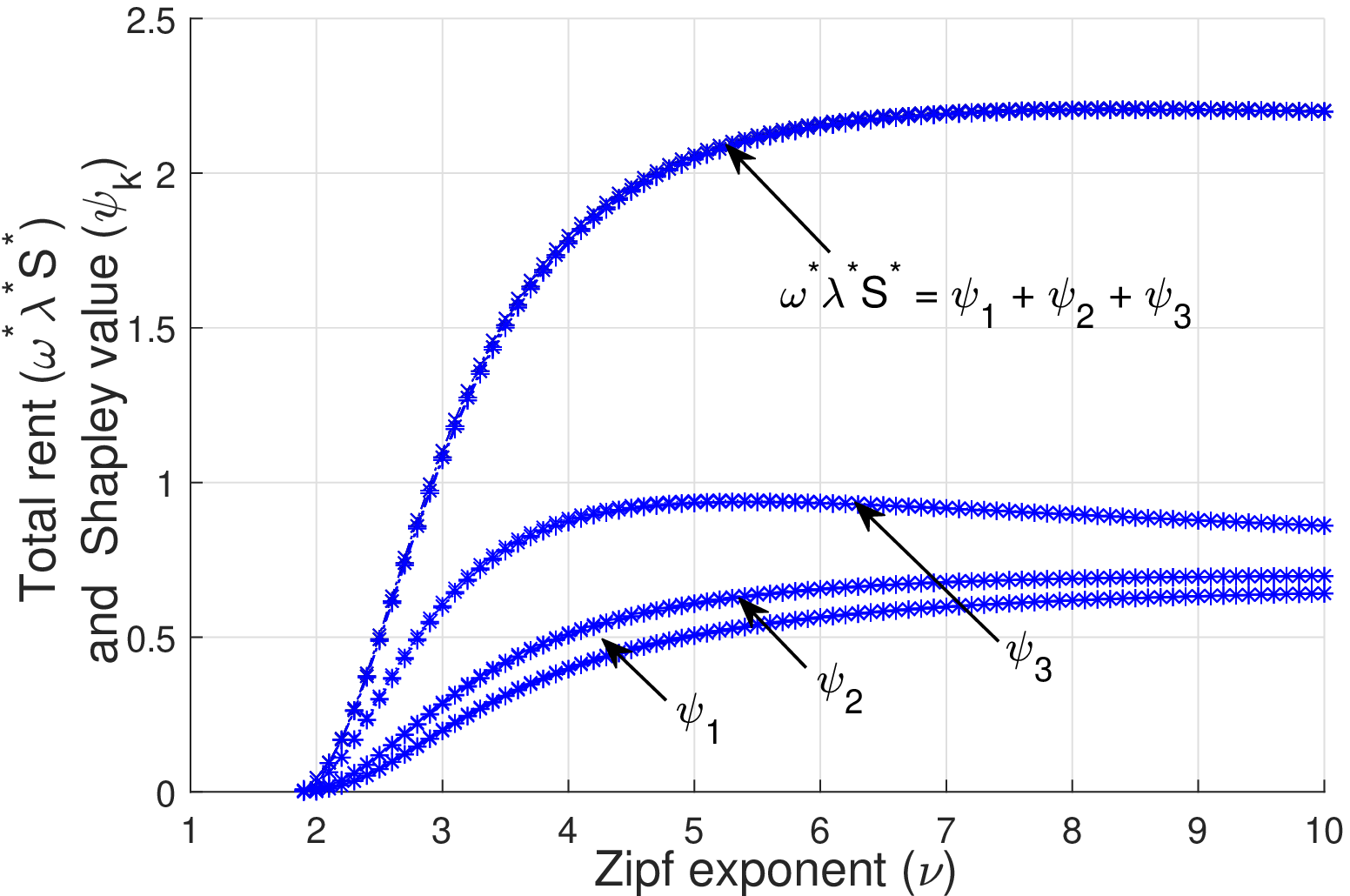}
        \caption{Total rent $\omega^*\lambda^*S^*$ and $\psi_k$ versus Zipf exponent ($\nu$) at Stackelberg equilibrium.}
        \label{fig:Totrent_Shapleyvalue_vs_Zipf}
     \end{center}
\end{minipage}
\hfill
\begin{minipage}{0.48\textwidth}
\begin{center}
	\includegraphics[width=\textwidth]{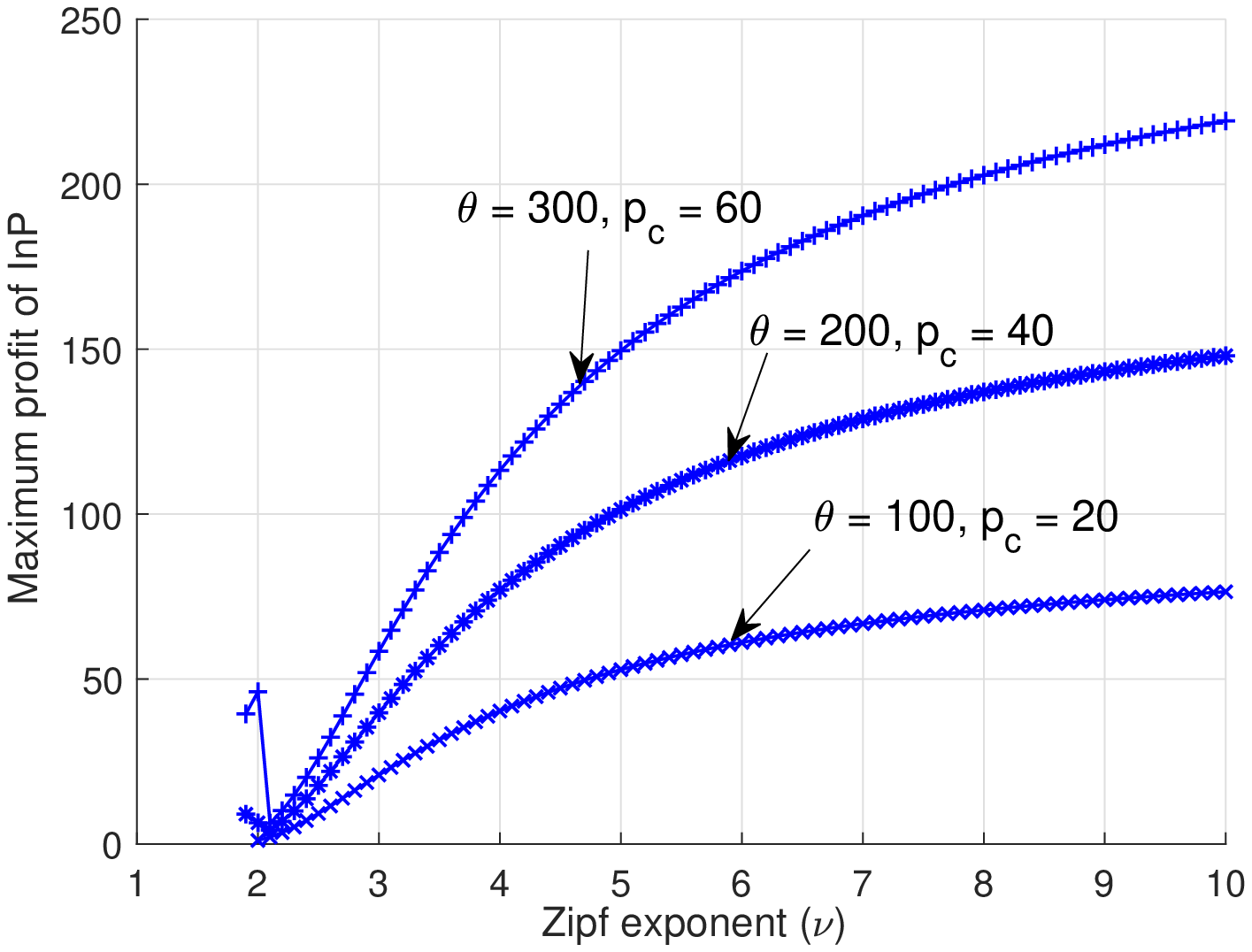}
        \caption{The maximum profit versus Zipf exponent ($\nu$) at Stackelberg equilibrium.}
        \label{fig:Maxprofit_vs_Zipf}
     \end{center}
\end{minipage}
\end{figure}

Fig.~\ref{fig:Totrent_Shapleyvalue_vs_Zipf} presents the total rent $\omega^*\lambda^*S^*$, which is the first term of the problem (Q0), and the  Shapley value of each MNO-$k$, $\psi_k$, versus the Zipf exponent, $\nu$. For a given value of $\nu$, we observe that $\psi_1$ is the lowest while $\psi_3$ is highest. This is because the MNO-$1$ has lowest bandwidth while MNO-$3$ has the highest bandwidth. As discussed in Section \ref{subsec:opt-strategy-MNO}, higher bandwidth, $W$, tends to increase the cache size $S^*$. Thus, the MNO-$3$ will try to buy the highest amount of infrastructure compared to MNO-$1$ and MNO-$2$. It can be seen that all three MNOs can divide the rent in a fair manner by using \textbf{Algorithm~\ref{alg:Airport_Runway_CostSharing}}, which is the Shapley value of their cooperative game. This can be confirmed by showing that $\omega^*\lambda^*S^* = \psi_1+\psi_2+\psi_3$ in the plot. The maximum profit of InP versus the $\nu$ is plotted in Fig.~\ref{fig:Maxprofit_vs_Zipf}, which is obtained by solving (Q0) using \textbf{Algorithm~\ref{alg:SGA_Algorithm}}. We observe that when $\nu$ increases, the maximum profit of InP is also increased. Also when $\theta$ and $p_c$ increase, the maximum profit of the InP is enhanced significantly.

The major observations from these numerical results are: (i) The Zipf exponent $\nu$ has a significant impact on the optimal strategy of an MNO. (ii) When the bandwidth $W$ changes, the optimal cache size $S^*$ also changes. However, changing $W$ does not affect the base station intensity $\lambda^*$. (iii) The price of areal power consumption $\theta$ and the circuit power $p_c$ affect the profit of the InP significantly.

\section{Conclusion}
\label{subsec:conclusion}
We have modeled and analyzed the performance of cache-enabled virtualized cellular networks by considering downlink SINR coverage probability and throughput, which are obtained based on stochastic geometry analysis. In these virtualized networks, the infrastructure, which consists of RAN and edge caching, is provided by the InP. With infrastructure sharing deployment, multiple MNOs can use the common infrastructure simultaneously.  For an MNO, the problem of minimization of the cost of sharing base stations and cache storage, subject to a probabilistic delay constraint,  is converted into a geometric program and a closed-form solution is obtained. Likewise, we have modeled the pricing problem for  sharing the caching infrastructure by using a Stackelberg game, where the InP is assumed to be the leader while the MNOs are assumed to be the followers. We have solved it via successive geometric programming. Lastly, sharing of the rent of infrastructure among the MNOs is done via Shapely value. We have observed that the Zipf exponent has significant effect on performance of MNOs, while the price of areal power consumption and the circuit power are the main parameters that affect the profit of the InP.

\appendix
The generalized harmonic number $H_{S,\nu}$ does not have a closed-form expression. Nevertheless, for analytical tractability, we can make an asymptotic approximation in terms of $S$ and $\nu$. To do so, we will relate the generalized harmonic number to the Hurwitz zeta function and then use the properties of Hurwitz zeta function. The Hurwitz zeta function, $\zeta(s,a)$, is defined as~\cite[Eqn 25.11.1]{DLMF2017}
\beq 
\zeta(s,a) = \sum_{n=0}^{\infty} \frac{1}{(n+a)^s}, 
\label{eqn:hurwitz-zeta-fun}
\eeq
 where $\Re(s) >1$ and $a \neq 0, -1, -2, \ldots$. The Hurwitz zeta function reduces to the Reimann zeta function when $a=1$,
$ \zeta(s,1) =\zeta(s)$, where $\zeta(s)$ is the Riemann zeta function. Also, harmonic sums can be expressed in terms of Hurwitz zeta function as~\cite[Eqn 25.11.4]{DLMF2017}
\beq 
\sum_{n=0}^{m-1} \frac{1}{(n+a)^{s}} = \zeta(s,a) - \zeta(s, a+m). 
\label{eqn:hurwitz-zeta-harmonic-sum-1}
\eeq

For our case, comparing (\ref{eqn:def-genearlized-harmonic-sum}) and (\ref{eqn:hurwitz-zeta-harmonic-sum-1}), we can express the generalized harmonic sum $H_{S,\nu}$ in terms of the Hurwitz zeta function as 
\beq 
H_{S,\nu} = \sum_{n=0}^{S-1}\frac{1}{(n+1)^\nu} = \zeta(\nu) - \zeta(\nu,S+1).
\label{eqn:hurwitz-zeta-harmonic-sum-2}
\eeq

Now, as $S \to \infty$, the asymptotic expansion of Hurwitz zeta function is given by~\cite[Eqn 25.11.43]{DLMF2017} 
\begin{align} 
\zeta(\nu,S+1) \sim & \frac{(S+1)^{1-\nu}}{\nu-1} + \frac{1}{2}(S+1)^{-\nu} + \sum_{k=1}^{\infty} \frac{B_{2k}}{(2k)!}(\nu)_{2k-1}(s+1)^{1-\nu-2k}, 
\label{eqn:hurwitz-zeta-asymptotic}
\end{align}
where $B_{2k}$ are Bernoulli numbers and $(\nu)_{2k-1} = \nu (\nu + 1) \cdots (\nu + 2k - 2)$ are Pochammer's symbol for rising factorial. Taking only the first dominant term from (\ref{eqn:hurwitz-zeta-asymptotic}) and substituting it in (\ref{eqn:hurwitz-zeta-harmonic-sum-2}), we obtain the asymptotic approximation for the generalized harmonic number as
\beq 
H_{S,\nu} \sim \zeta(\nu) - \frac{(S+1)^{1-\nu}}{\nu -1}. 
\label{eqn:harmonic-sum-asymptotic}
\eeq

Thus, from the above arguments, using (\ref{eqn:hit-probability}) and (\ref{eqn:harmonic-sum-asymptotic}), we have the desired lemma.

\bibliographystyle{IEEE}

\end{document}